# Unified University Inventory System (UUIS)

# Software Requirements Specifications

## Authors

Ahmed Daoudi
David Zerkler
Gay Hazan
Isabelle Toutant
Mariano Diaz
René Toutant
Virginia Cook
William Nzoukou
Yassine Amaiche

## Revision History

| Revision | Date | Comments |
|---|---|---|
| 1.0 | 2010-03-02 | First draft |
| 1.1 | 2010-03-04 | Sections 1.1, 1.4, 2.2 and 2.3 added |
| 1.2 | 2010-03-06 | Sections 3.2.6 and 3.2.7 added |
| 1.3 | 2010-03-10 | Sections 3.2.1, 3.2.3, appendix A,B,C and D added |
| 2.0 | 2010-03-22 | Merged Problem Reporting/Tracking and Requests/Approvals/Work Orders into a single System Feature called Requests. Update Physical Asset Management |
| 2.1 | 2010-03-22 | Removed class diagrams, ER diagrams and data dictionary |
| 2.2 | 2010-03-22 | Sections 3.2.4 added; Removed Cocomo and Log sheets |
| 2.3 | 2010-03-23 | Added Non-Functional Requirements from A2 |
| 2.4 | 2010-04-21 | Added The last missing Parts Except User Characteristics |
| 2.5 | 2010-04-25 | Final edit |
| 2.6 | 2010-05-01 | Added Bulk Entry Use Case & FR. |

# Table of Content

# List of Figures

# List of Tables

# 1. Introduction

## 1.1 Purpose and Scope

This Software Requirement Specification (SRS) provides the detailed requirements for the Unified University Inventory System (UUIS) produced for the Imaginary University of Arctica (IUfA). This document is intended to be used primarily by the software designers during the design phase, but will also be used extensively during the testing phase.

The UUIS will integrate and organize the existing inventory systems of three faculties in the IUfA. While the system will support the faculties existing legacy inventory codes and barcodes, emphasis will be placed on the new unified barcode system.

The system's primary function will be to manage all types of physical assets, software licenses, and location contents & custodians. Managing, in this context, implies keeping a record of the current location, status, description, and ownership of the inventoried item, as well as being able to search for and display any system information.

Physical assets include all manors of furniture and equipment, whether movable (desks, chairs, computers, etc) or immovable (projectors fixed to a room). Software licenses will be managed, including who has access to which license, and where (on which computer) the software is installed. Locations are all the places in which things are stored, which includes the various types of rooms in the university, as well as lockers, and cabinet drawers. Both the contents of the location and the location's owner are stored.

The secondary role of the software is to create a system of work orders, requests and problem tickets to support the proper movement and allocation of the inventoried items. UUIS will maintain a list of individual user permissions and access rights that prevent unauthorized access to potentially sensitive information.

The main advantage of the UUIS will be the use of all university inventory information in a clear, concise and consistent format. This will allow a smaller learning curve for students, faculty and employees in manipulating information. UUIS will be especially useful for any administrator making decisions based on the inventory of the entire university, and avoiding the use of multiple individual faculty systems.

## 1.2 Acronyms

| | |
|---|---|
| E/R | Entity Relationship |
| HTTP | Hypertext Transfer Protocol |
| IDE | Integrated Development Environment |
| IUfA | Imaginary University of Arctica |
| RDBMS | Relational database management system |
| SRS | Software Requirement Specification |
| TBD | To Be Determined |
| UUIS | Unified University Inventory System |

## 1.3 References

[IEEE, 1998] *std 830-1998: IEEE Recommended Practice for Software Requirements Specifications* by Institute of Electrical and Electronics Engineers, 1998.

## 1.4 Overview

Section 2 describes the needs and constraints on the UUIS software as well as the hardware systems it will run on and interface with.

Section 3 provides a detailed description of all the UUIS abilities and the requirements necessary to correctly enable those abilities.

This SRS is designed with real-world objects in mind, and relies heavily on use cases to explain the various functions associated with those objects. A list of specific functional requirements follows each section in order to delineate exactly what needs to be done by the programmer. These functional requirements allow an adequate organization and simplification of the testing phase.

Section 3 concludes with a detailed description of various physical, logical, and security constraints. This takes the form of a list of non-functional requirements.

# 2. Overall Description

This section of the SRS provides a background for the requirements detailed in section 3.

## 2.1 Product Perspective

UUIS is a web-based client application that can be accessed via any web browsers over a network. It is a 3-tier solution in which the user interface, the business logic and the data management are developed and maintained as independent modules.

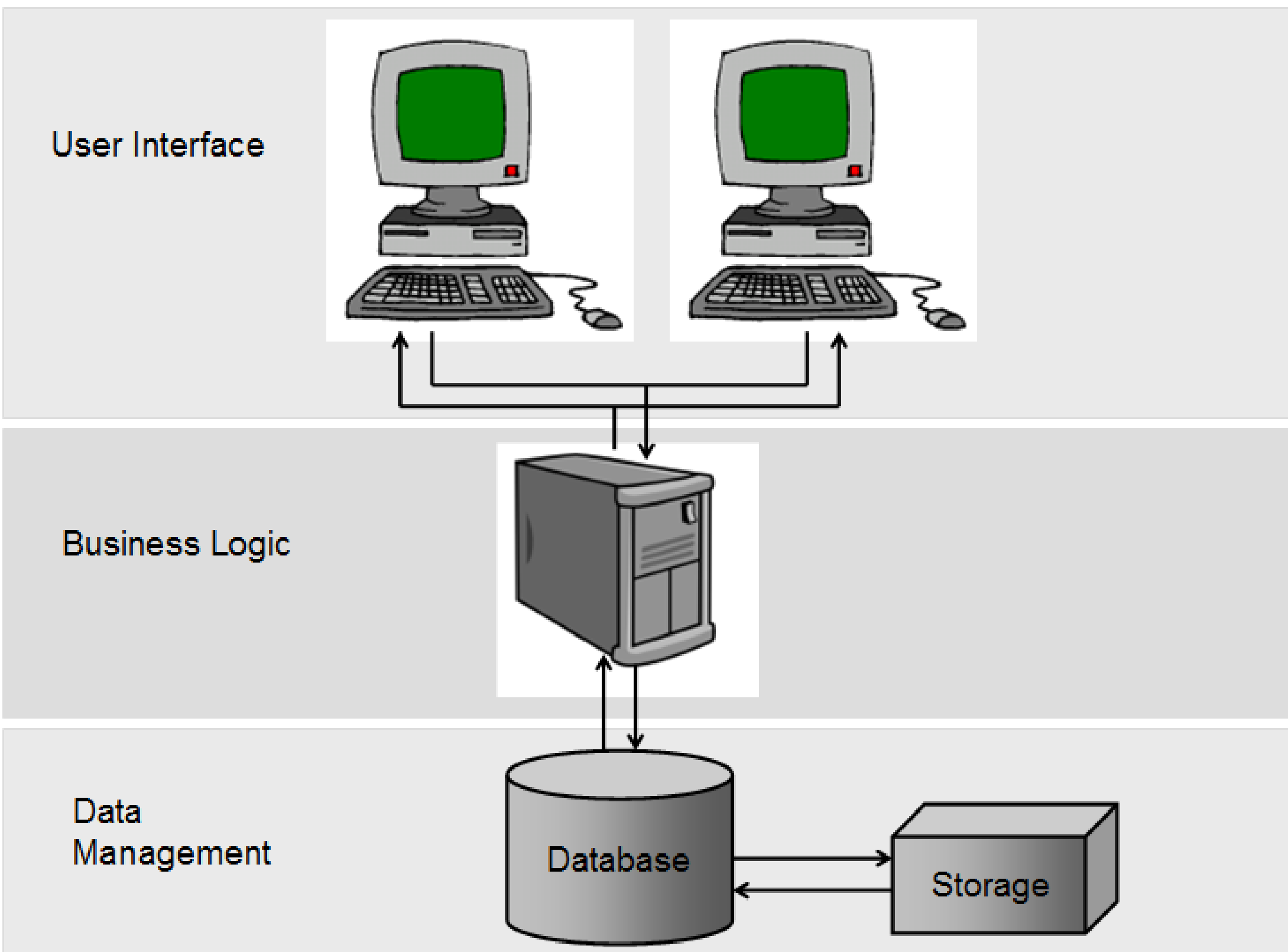

**Figure 1 -** UUIS Architecture

*1. User Interface*
Top layer of the application, the interface main function is to display information received from other tiers in a user-friendly format.

*2. Business Logic*
This middle layer coordinates the application by handling and processing information exchanges between the database and the user interface, and by making logical decisions and performing calculations.

*3. Data Management*
Bottom layer of the application, the data management consists of a database server that stores all system's data. It keeps data independent from application servers or business logic.

### 2.1.1 System Interfaces

UUIS is a web-based client interface that will interact with the database server using TCP/IP connection.

### 2.1.2 User Interfaces

The user will interact with the system through a web interface supported by any computer running one of the following operating systems: Microsoft Windows, UNIX, Linux, or Apple Macintosh. The interface will run properly provided the computer has a web-browser installed (Microsoft Internet Explorer, Mozilla Firefox, etc.) and internet connection is available.

All elements of the interface will be displayed in English. However, if required, the design will support the addition of other languages.

### 2.1.3 Hardware Interfaces

No specific hardware requirements.

### 2.1.4 Memory Constraints

- Minimum memory required for Eclipse IDE installation: 189 MB;
- Minimum memory required for Tomcat installation: 9 MB;
- Minimum memory required for MySQL installation: 104.9 MB;
- Size and quantity of stored data; and
- spec107 hard drive specifications.

### 2.1.5 Software Interfaces

- Apache Tomcat version 6.0.24, an open source Java web-server;
- MySQL version 5.1.43, a relational database management system (RDBMS) that runs as a server providing multi-user access to a given database;
- Java version 6; and
- Eclipse release 3.5.1 as a software integrated development environment (IDE).

### 2.1.6 Communication Interfaces

UUIS uses Hypertext Transfer Protocol (HTTP) as client-server communication protocol. The interaction with the database server will be made using TCP/IP connection.

### 2.1.7 Operations

Backups for the Database of the system will be performed on a daily basis.

## 2.2 Product functions

The UUIS is a tool to organize the inventory needs of the IUfA. This includes:

- Maintaining a list of all inventory assets, including legacy and/or universal barcodes and multiple descriptive attributes per inventory item;
- Managing the current location of all inventory assets, including off-campus locations (e.g. professor's home, in the field, on a job-site, etc.);
- Managing a system of requests, approvals and confirmations to create work-orders and detail the movement of inventory assets;
- Managing detailed software-licensing agreements, including installation locations;
- Managing the stewardship and contents of locations (e.g. offices, lockers, labs, lab seats, desk drawers, etc.);
- Maintaining a rigid set of User Permissions to ensure access to user appropriate data ONLY (searching and/or editing);
- Creating a variety of reports including detailed log of system use; and
- Reporting errors or problems such as:
    a. Broken furniture or equipment
    b. Missing furniture or equipment
    c. Expiring Warrantees
    d. Room overcapacity

## 2.3 User characteristics

The UUIS is designed to be usable by individuals with a minimum level of literacy related to the use of information technology and computers. The end users of the system should have a functional knowledge of the usage of a web browser, such as opening a web browser, navigating through WebPages, filling-in and submitting forms, and the general interaction with a dynamic website.
The users should also be able to read English and understand the text to navigate through the system. Since the system does not provide any accessibility features, it is not suitable for use by people with visual disabilities.

The following tables list the main type of users and stakeholders and give a brief description of their usual needs.

**Table 1** – UUIS IUfA users' description

| ROLE | USERS | DESCRIPTION |
|---|---|---|
| 0 | Student | Allowed to request or report problems |
| 1 | Teacher | Make requests and access to department database |
| | Department Administrator | Make requests and access to department database |
| | IT group member | Make requests and access to department database |
| 2 | Faculty Administrator | Access to faculty database |
| 3 | University Administrator | Complete access to database |

## 2.4 Constraints

The UUIS is a dynamic web application and it will be implemented using the PHP language. Therefore a web server (preferably Apache 2 and above) and a PHP interpreter (version 5.2 and above) are required.

## 2.5 Assumptions and Dependencies

The UUIS will be developed using the PHP language. Therefore, it is assumed that the system on which the UUIS will be deployed will have an Apache web server, a PHP interpreter and a MySQL Database up and running.

## 2.6 Apportioning of requirements

Future versions of the UUIS website will support multiple languages. Priority will be the addition of French. Other languages will be added on an as-required basis.

Bar-coding technology might be supported in future versions to add or search for physical assets in the system. Bar-coding is a technology that has proven its reliability worldwide. Its main advantage is the automation of some tasks that are manually done now by a human operator.

# 3. Specific Requirements

## 3.1 Descriptive Requirements

### 3.1.1 User Interfaces

Appendix A provides a detailed description of the UUIS user interfaces.

### 3.1.2 Hardware, Software and Communications Interfaces

The software will be installed on three Dell PowerEdge T710 Tower Server machines that will be interconnected to each other. Different modules will be installed on each server. A fourth server will be used as a DBMS dedicated machine to which all the other three servers will have to connect in order to execute their queries and get results back. The software needed is a popular browser installed on client machines. A web server and a PHP interpreter will reside on three servers, and a RDBMS will reside on the fourth server (MySQL). The main communication protocol will be the one used for internet all over the world, TCP/IP. Requests sent will be HTTP formatted.

## 3.2 System Features

The following figure illustrates the partitioning of the IUfA UUIS from a use cases and additional functions standpoint.

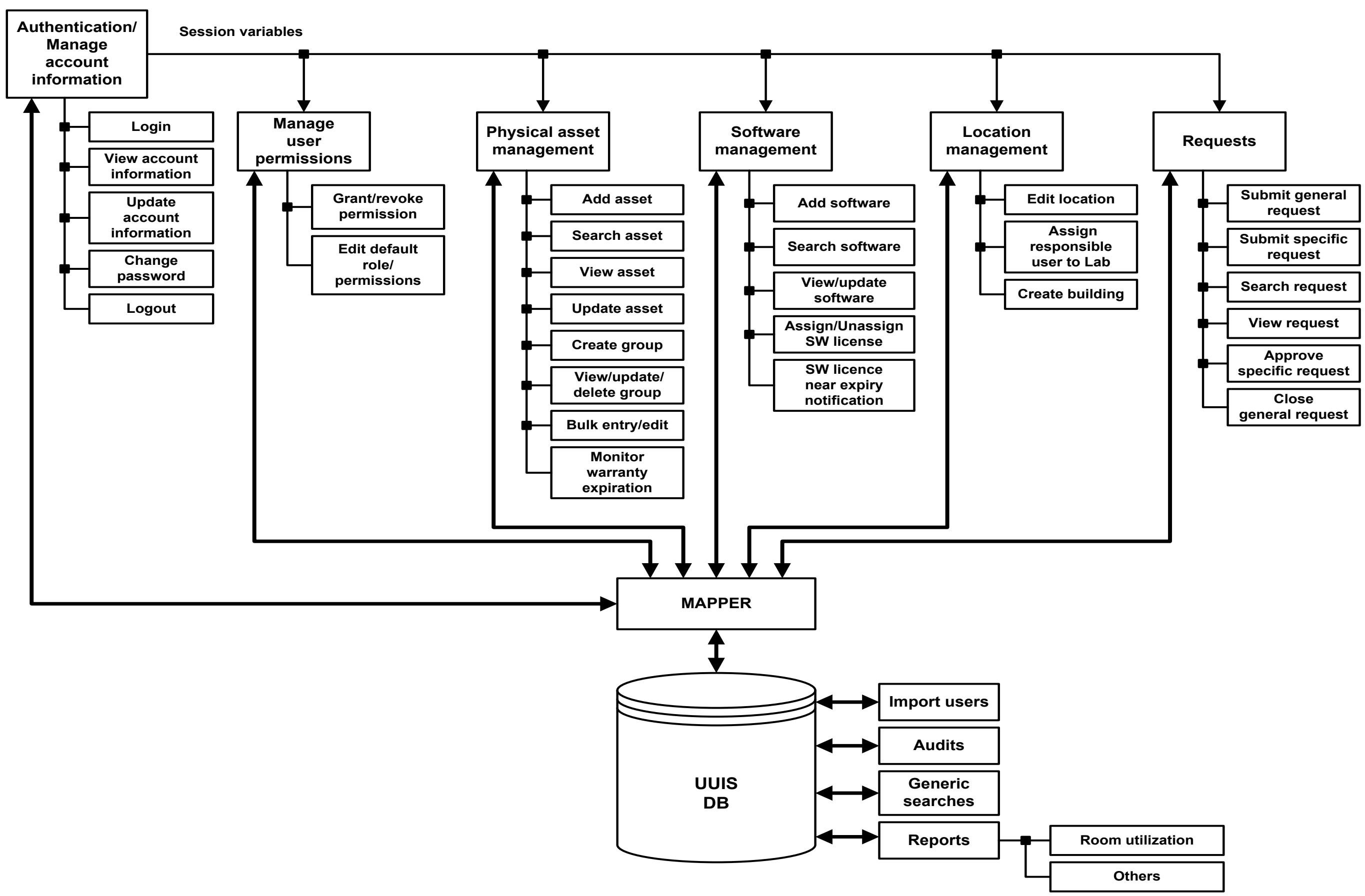

**Figure 2** – System Features – Bloc Diagram

## 3.2.1 Authentication/Manage Account Information

The Authentication/Manage Account Information feature is required in the UUIS application to ensure only legitimate users are logged on to the system and can access information. Authentication determines also what kind of access level a user is entitled to.

### 3.2.1.1 Use Case Model

The following figure provides the Authentication/Manage Account Information use case diagram.

**Figure 3 -** Authentication/Manage Account Information – Use case diagram

The following paragraphs provide the detailed description for each use case.

*3.2.1.1.1 Login*

**Role:** General users (Role 0);
**Pre-condition:** Unauthenticated session;
**Steps:**

1. The user selects the *Login* button accessible from the *IUfA UUIS Home page*;
2. The system displays the *Login* page;
3. The user provides username, password, and string of displayed image (CAPTCHA);
4. The user selects the *Submit* button;
5. The system, if the provided information is incomplete or erroneous, displays an error message to the user:
    a. The user selects the *Return* button; and

b. The system displays the *Login* page with blank fields;
6. The system fetch user permission level and displays IUfA UUIS main page with menus corresponding to user permission level;
7. The system:
    a. Creates a new log record to the database;
    b. Assigns the log entry a unique identification number (logID);
    c. Registers the user ID, and login date;

**Post-conditions:**
1. Authenticated session; and
2. A new log entry is stored in the database;

**Exceptions:**
1. The user cancels the login by selecting the *Cancel* button from the *Login* page:
    a. The system displays the *IUfA Home* page; and
    b. The system does not store any updated information;
2. A database error is encountered:
    a. The system displays an error message;
3. Invalid username, password and/or string (CAPTCHA):
    a. The system displays an error message.

*3.2.1.1.2 View Account Information*

**Role:** General users (Role 0);
**Pre-condition:** Authenticated session;
**Steps:**
1. The user selects the *View account* button under the *MyProfile* tab (accessible from all system pages);
2. The system displays the *View Account Information* page with:
    a. Username
    b. First name;
    c. Last name;
    d. E-mail;
    e. Role;

**Post-conditions:** The system displays the *View Account Information* page;
**Exceptions:**
1. A database error is encountered:
    a. The system displays an error message.

*3.2.1.1.3 Update Account Information*

**Role:** General users (Role 0);
**Pre-condition:** Authenticated session;

**Steps:**

1. The user selects the *Update account* button under the *MyProfile* tab (accessible from all system pages);
2. The system displays the *Update Account Information* page with:
    a. Un-editable fields:
        i. Username;
        ii. Role;
    b. Editable fields (all mandatory);
        i. First name;
        ii. Last name;
        iii. E-mail;
3. The user edits/updates the above fields as applicable;
4. The user selects the *Submit* button;
5. The system, if any of the above fields are incomplete, displays a message indicating the missing information and request the user to provide it:
    a. The user selects the *Return* button; and
    b. The system displays the *Update Account Information* page with current information;
6. The system displays a message with the user provided information and requesting the user to confirm the submission:
    a. The user selects the *Confirm* button; or
    b. The user selects the *Return* button;
        i. The system displays the *Update Account Information* page with current information;
7. The system updates the User database entry with the applicable changes;

**Post-conditions:**

1. The system displays the *View Account Information* page with updated information; and
2. The updated user information is stored in the database;

**Exceptions:**

1. The user cancels the update by selecting the Cancel button from the *Update Account Information* page:
    a. The system displays the *View Account Information* page; and
    b. The system does not store any updated information;
2. A database error is encountered:
    b. The system displays an error message.

*3.2.1.1.4 Change Password*

**Role:** General users (Role 0);
**Pre-condition:** Authenticated session;
**Steps:**

1. The user selects the *Change Password* button under the *MyProfile* tab (accessible from all system pages);
2. The system displays the *Change Password* page that requests the user to confirm old password once, and to enter new password twice;
3. The user enters the appropriate information and select the *Submit* button;
4. The system, if the provided information is incomplete or erroneous, displays an error message to the user:
    a. The user selects the *Return* button; and
    b. The system displays the *Change Password* page with blank fields;
5. The system:
    a. Validates the old and new password;
    b. Updates the User database entry with the new password;
    c. Displays a confirmation message to the user that password has been successfully changed;
        i. The user selects the *Return* button;
    d. An e-mail is sent to the user with password change confirmation;

**Post-conditions:**

1. The system displays the *Password Successfully Changed* page; and
2. The updated user password is stored in the database;

**Exceptions:**

1. The user cancels the update by selecting the Cancel button from the *Change Password* page:
    a. The system displays the *Change Password* page with blank fields; and
    b. The system does not store any updated information;
2. Invalid old or new passwords are entered:
    a. The system displays the an error message;
3. A database error is encountered:
    a. The system displays an error message.

*3.2.1.1.5 Logout*

**Role:** General users (Role 0);
**Pre-condition:** Authenticated session;
**Steps:**

1. The user selects the *Logout* button accessible from all system pages;
2. The system displays a message requesting the user to confirm the logout:
    a. The user selects the *Confirm* button; or
    b. The user selects the *Return* button;
        i. The system displays the original system page from which the *Logout* button was accessed);
3. The system updates the log database entry with the logout date;
4. The system displays the *IUfA UUIS Home* page.

**Post-conditions:**

1. Unauthenticated session; and
2. The logout date is stored in the database;

**Exceptions:**

1. A database error is encountered:
   a. The system displays an error message.

### 3.2.1.2 Functional Requirements

*3.2.1.2.1 Login*

**[F.Req.1]** The system shall show the user two username and password boxes

**[F.Req.2]** The system shall hide the password stroke keys from the user

*3.2.1.2.2 View Account Information*

**[F.Req.3]** The system shall show the user his level of access

**[F.Req.4]** The system shall show the user information on his last login session.

**[F.Req.5]** The system shall show the user his name and e-mail account

*3.2.1.2.3 Update Account Information*

**[F.Req.6]** The system shall allow the user to change his e-mail address

**[F.Req.7]** The system shall allow the user to change his home address

**[F.Req.8]** The system shall allow the user to change his phone number

*3.2.1.2.4 Change Password*

**[F.Req.9]** The system shall ask for the actual and the new password

**[F.Req.10]** The system shall ask for confirmation of the new password

**[F.Req.11]** The system shall send an e-mail to the user notifying him of password change.

*3.2.1.2.5 Logout*

**[F.Req.12]** The system shall show a logout Button on all the pages

**[F.Req.13]** The system shall close the user session after user logs out.

## 3.2.2 Manage User Permissions

Admin Users (Role 1 and above) of the IUfA UUIS have the capability to grant/revoke permissions. Following the request to grant or revoke permission from a Supervisor, the Admin User of each level will be able to:

1. The system shall allow the importing of data already available from a legacy data source.
2. Grant/Revoke the requested permission/s to a User;
3. Edit Default permissions to a given Role.
4. Add new Role.

### 3.2.2.1 Domain Model

The following figure illustrates the domain model applicable to Manage Permissions in the context of the IUfA UUIS.

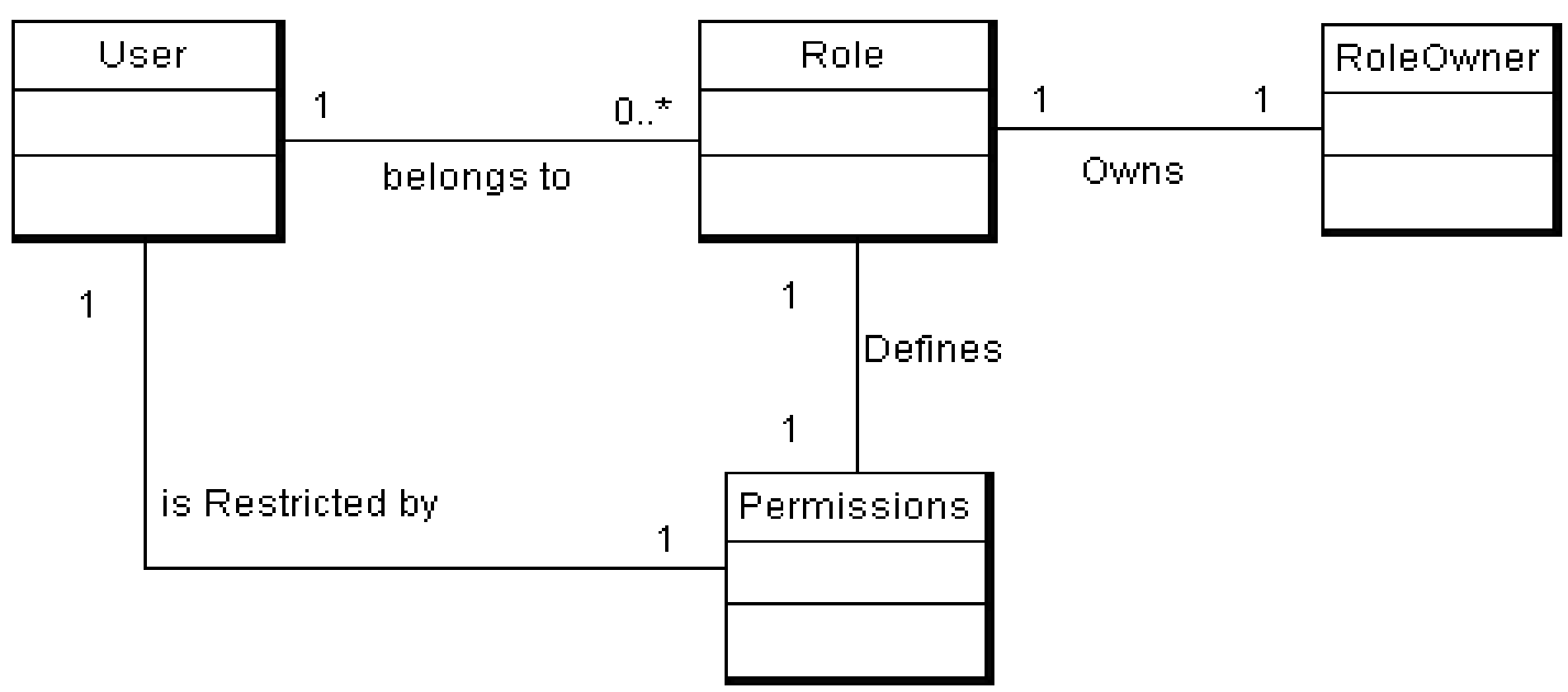

**Figure 4 –** Manage User Permission – Domain model

### 3.2.2.2 Use Case Model

The following figure provides the Manage User Permissions use case diagram.

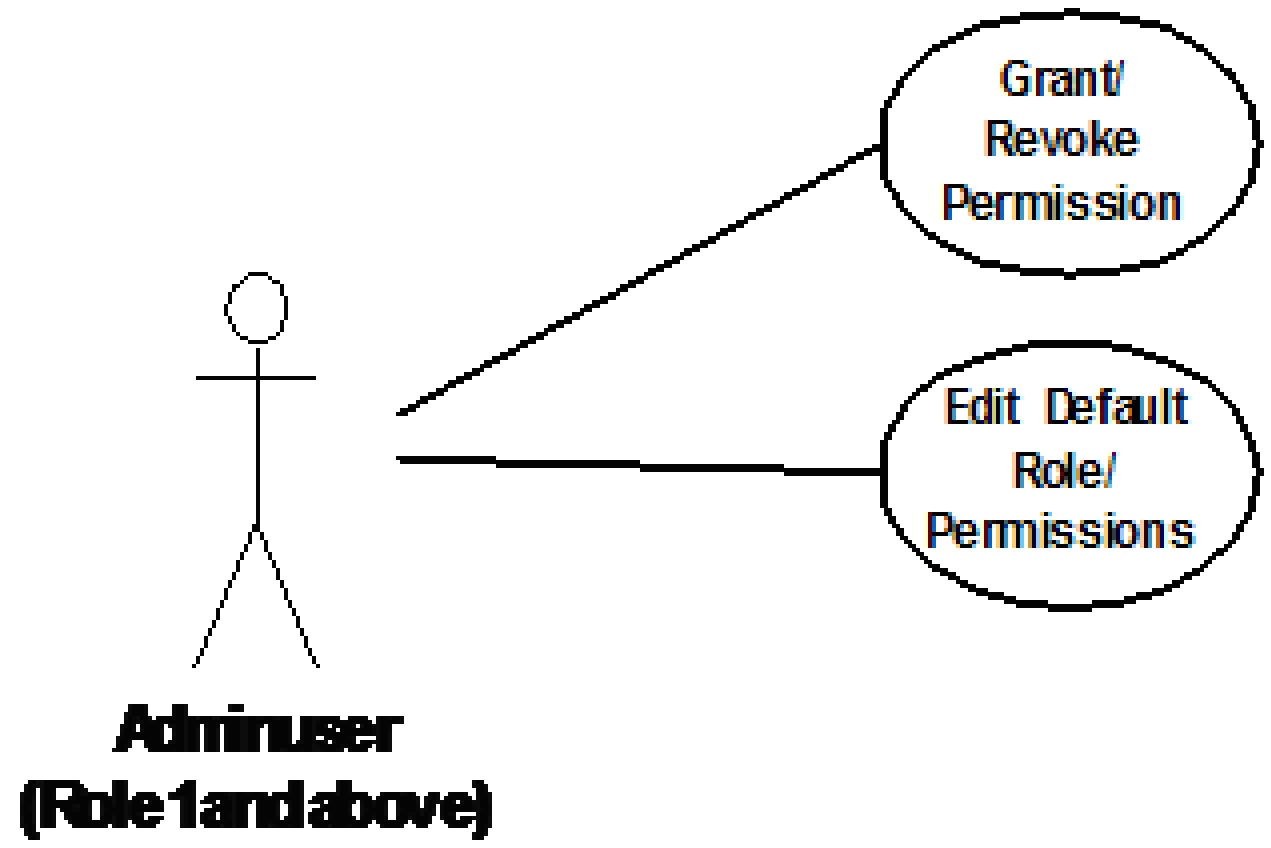

**Figure 5 –** Manage User Permission – Use case diagram

The following paragraphs provide the detailed description for each use case.

*3.2.2.2.1 Grant/Revoke Permission*

**Role:** Admin Users (Role 1 and above);
**Pre-condition:** Authenticated session, Admin privileges;
**Steps:**

1. Admin selects the *Security* option under the *System Administration* tab (accessible from all system pages);
2. The system displays the *Security Administration* page;
3. The Admin selects the *User Permission* Tab;
4. The Admin searches for target user;
5. The system displays the complete set of permissions, and whether they are Granted/Revoked for the selected user;
6. The admin selects permission to Grant/Revoke;

Repeat step 3-5 until finish;

7. The user selects the *Submit* button;
8. The system logs the transaction;

**Post-conditions:**

1. The system displays the *Security Administration* page;
2. A new set of permissions is Granted/Revoked to a given set of user/s;

**Exceptions:**

1. At any time system fails:

a. The system displays an error message.
2. At any time, the Admin cancels by selecting the Cancel button:
   a. The system displays the *Security Administration* page; and
   b. The system does not log transaction associated with the cancelled request;

The system need to ensure a continuous up time all the time. If an unforeseen event happens that causes the main servers to go down, backups in another server need to be still available for access. Whenever a transaction happens in the main server like write, the secondary server needs to perform the same action.
The set of permissions in the system is restricted to the ones formally defined. There is no need for customized permissions since roles and status of everyone in the university is already known in advance.

*3.2.2.2.2 Edit Default Role Permissions*

**Role:** Admin Users (Role 1 and above);
**Pre-condition:** Authenticated session, Admin privileges;
**Steps:**

1. Admin selects the *Security* option under the *System Administration* tab (accessible from all system pages);
2. The system displays the *Security Administration* page;
3. The Admin selects the *Role Permission* Tab;
4. The Admin searches for a *Role* to edit;
5. The system displays the complete set of permissions, and whether they are Granted/Revoked for the selected Role;
6. The admin selects permission to Grant/Revoke;

   Repeat step 3-5 until finish;

7. The user selects the *Submit* button;
8. The system logs the transaction;

**Post-condition:**

1. The system displays the Security Administration page;
2. A new set of permissions is Granted/Revoked to a given set of user/s;

**Exceptions:**

1. At any time system fails:
   a. The system displays an error message.
2. At any time, the Admin cancels by selecting the Cancel button:
   a. The system displays the *Security Administration* page; and
   b. The system does not log transaction associated with the cancelled request;

#### 3.2.2.3 Functional Requirements

*3.2.2.3.1 Grant/Revoke Permissions*

**[F.Req.1]** The system shall allow Admins to search for users.

**[F.Req.2]** The system shall allow Admins to clear search criteria

**[F.Req.3]** The system shall allow Admins of a given Role, to Grant/Revoke permissions to users belonging to that given Role.

**[F.Req.4]** The system shall log every Transaction involving Permissions for further Audit.

**[F.Req.5]** The system shall allow a Admins to cancel a transaction.

**[F.Req.6]** The system shall allow Admins to Grant/Revoke Admin Privileges to a given user.

**[F.Req.7]** The system shall, upon successfully completing the report problem operation, issue an e-mail notification to the user and the Problem tracking coordinator including unique identification number and date reported.

*3.2.5.3.1 Edit Default Role Permissions*

**[F.Req.8]** The system shall allow Admins to search for Roles.

**[F.Req.9]** The system shall allow Admins of a given Role, to edit the default set of permissions of that given Role

**[F.Req.10]** The system shall log every Transaction involving Permissions for further Audit.

**[F.Req.11]** The system shall allow a Admins to cancel a transaction.

*3.2.5.3.3 Add New Role*

**[F.Req.12]** The system shall allow Admins to search for Roles

**[F.Req.13]** The system shall allow Admins to set the default set of permissions for a new Role.

**[F.Req.14]** The system shall log every Transaction involving Permissions for further Audit.

**[F.Req.15]** The system shall allow a Admins to cancel a transaction.

*3.2.5.3.4 General/Miscellaneous*

**[F.Req.16]** The system shall allow Admins to print a general user permissions report.

**[F.Req.17]** The system shall allow Admins to print a user permissions report by Roles.

**[F.Req.18]** The system shall allow Admins to import users from an existing system.

### 3.2.3 Requests

All users of the IUfA UUIS have the capability to submit requests. The IUfA UUIS supports two types of requests, general and specific.

General requests are broken into two categories:

a. Administrative request; and
b. Technical request.

General administrative requests are used to report problems such as:

a. Malfunctioning physical asset;
b. Broken physical asset; and
c. Lost physical asset.

General technical requests are used to report problems involving IT such as:

a. Malfunctioning computer component;
b. Broken computer component;
c. Malfunctioning network;
d. Malfunctioning software;
e. Request software installation; and
f. Request a modification to one's user profile.

General administrative requests will be monitored by a designated individual for each level (based on the requester level). General technical requests will be monitored by a Role 3 designated individual. The designated individual will take the appropriate action and close the request (and provide a closure note).

The IUfA UUIS supports the five categories of specific requests listed below:

a. Move asset to location;
b. Move group to location;
c. Assign equipment to user;
d. Assign storage compartment to user; and
e. Assign group to user.

Specific requests that involve an intra-faculty transaction only (e.g. asset from a faculty assigned to a user of that faculty) require approval at the faculty level. Specific requests that involve a faculty to faculty transaction (e.g. asset from a faculty assigned to a user from another faculty) require approval at the university level.

Users can browse all requests and their status at their level and below for their department. Level 0 users can only browse their own requests.

### 3.2.3.1 Activity Diagram

The following figure provides the activity diagram applicable to the management of requests in the context of the IUfA UUIS.

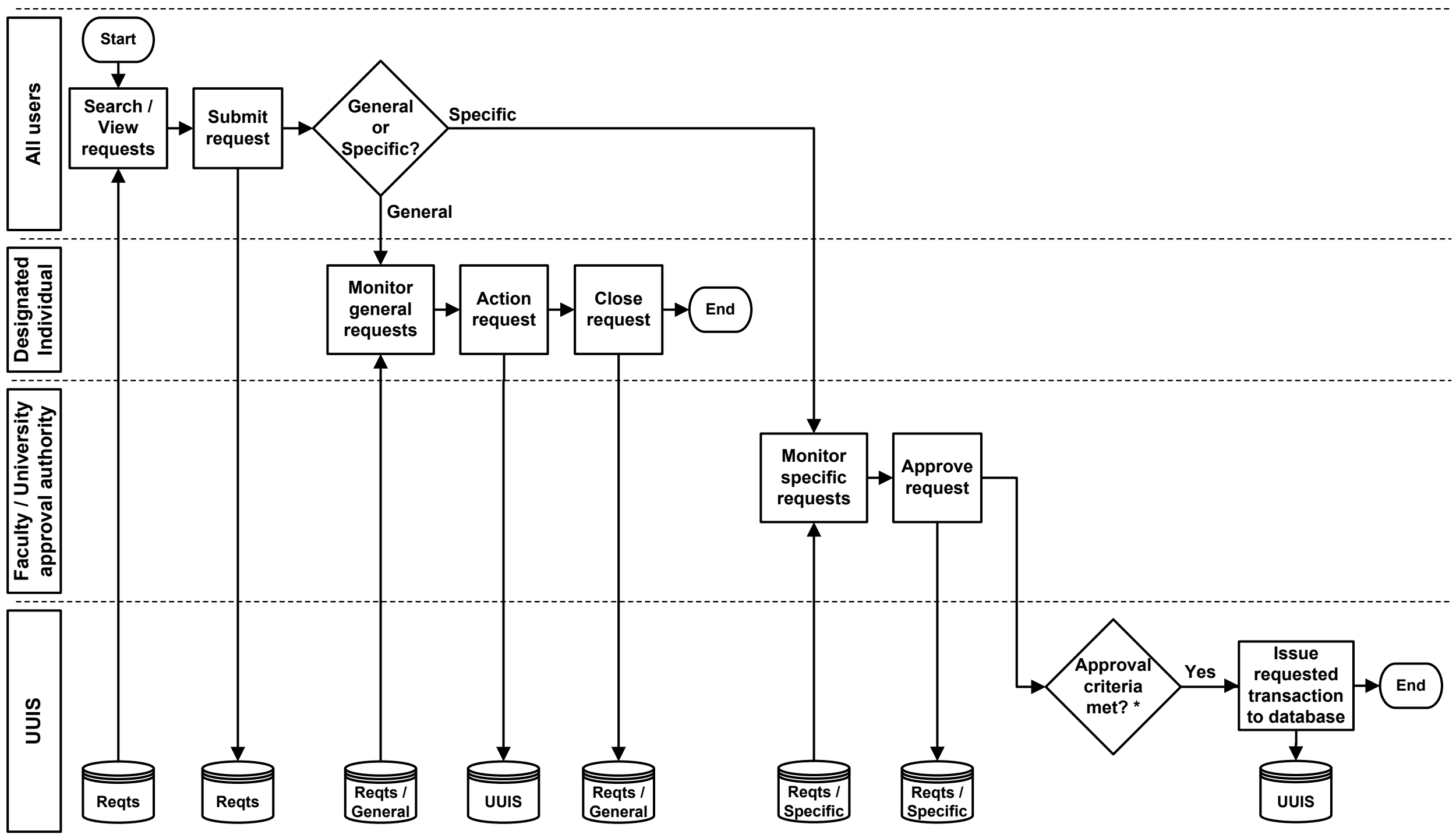

* Based on request (intra-faculty versus faculty-to-faculty) and user's role and faculty
Note: Specific requests that meet approval criteria upon submission are immediately approved and issued to the database

**Figure 6 –** Requests – Activity diagram

### 3.2.3.2 Use Case Model

The following figure provides the Requests use case diagram.

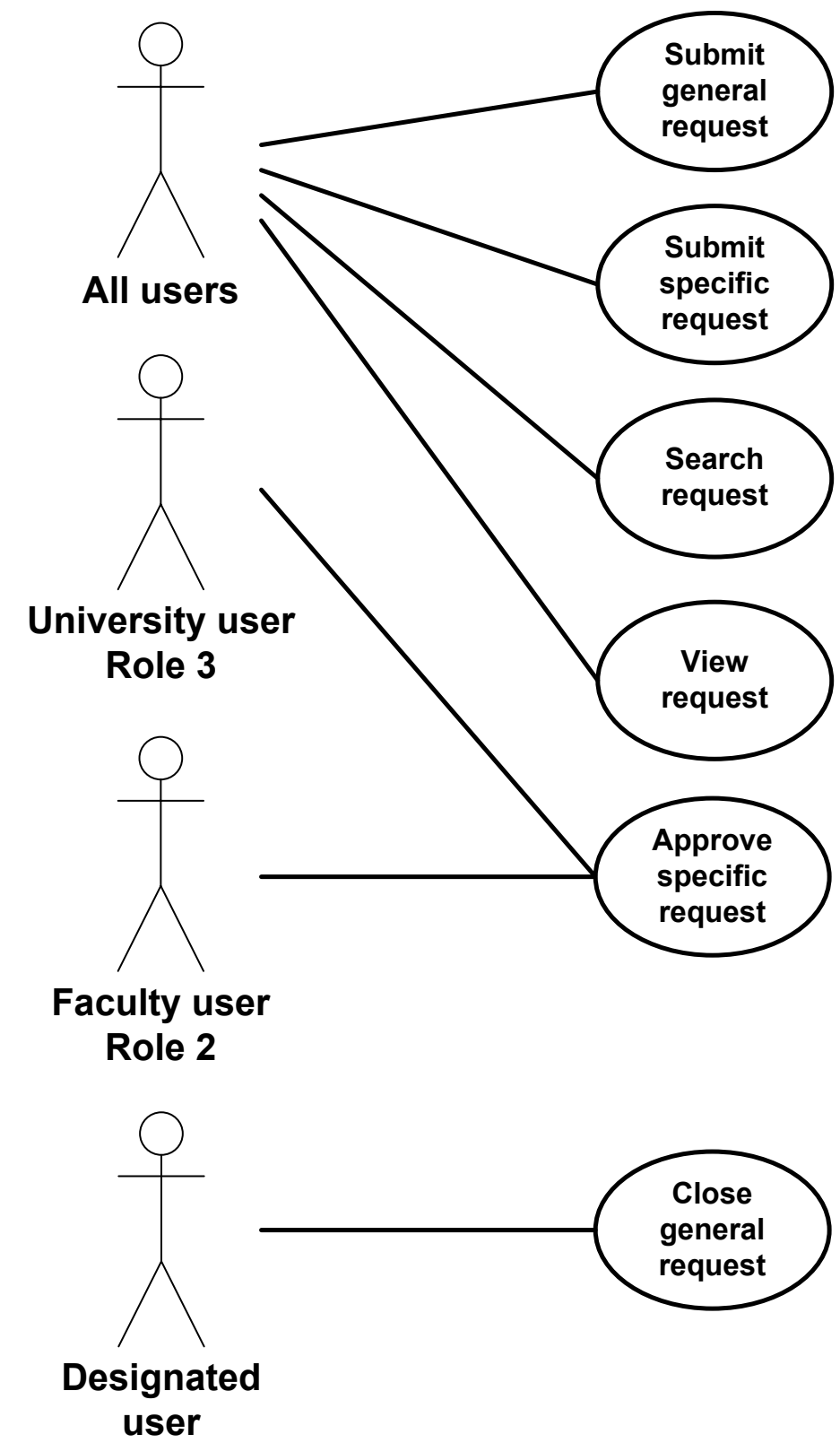

**Figure 7 –** Requests – Use case diagram

The following paragraphs provide the detailed description for each use case.

*3.2.3.2.1 Submit General Request*

**Role:** All users;
**Pre-condition:** Authenticated session;
**Steps:**

1. The user selects the *Submit general request* button under the *Requests* tab (accessible from all system pages);
2. The system displays the *Submit general request* page;
3. The user provides the following information:
    a. Category – radio button with choices *General administrative* or *General technical*; and
    b. Description;
4. The user selects the *Submit* button;

5. The system, if a description is not provided, displays a message and requests the user to provide it:
    a. The user selects the *Return* button; and
    b. The system displays the *Submit general request* page with current information;
6. The system:
    a. Assigns the request a unique identification number (request ID);
    b. Sets the request status to *Pending*;
    c. Adds the request to the database including request ID, category, description, status, originator username, faculty, department and level;
    d. Displays a message to the user with the request ID;
        i. The user selects the *Return* button;

**Post-conditions:**

1. The system displays the system page from which the *Requests* tab was accessed; and
2. The new request is stored in the database;

**Exceptions:**

1. The user cancels the request by selecting the *Cancel* button from the *Submit general request* page:
    a. The system displays the original system page (from which the *Requests* tab was accessed); and
    b. The system does not store any information associated with the cancelled request;
2. A database error is encountered:
    a. The system displays an error message.

*3.2.3.2.2 Submit Specific Request*

**Role:** All users;
**Pre-condition:** Authenticated session;
**Steps:**

1. The user selects the *Submit specific request* button under the *Requests* tab (accessible from all system pages);
2. The system displays the *Submit specific request* page;
3. The user selects one of the following choices (radio button) and provides the applicable (mandatory) information:
    a. *Move asset to location*:
        i. Barcode; and
        ii. Location;
    b. *Move group to location*:
        i. Group ID; and
        ii. Location;
    c. *Assign equipment to user*:

i. Barcode; and
            ii. Username;
        d. *Assign storage compartment to user*:
            i. Barcode;
            ii. Compartment number; and
            iii. Username;
        e. *Assign group to user*:
            i. Group ID; and
            ii. Username;
4. The user selects the *Submit* button;
5. The system, if one or more mandatory field is missing, displays a message indicating the missing information and requesting the user to provide it:
    a. The user selects the *Return* button; and
    b. The system displays the *Submit specific request* page with current information;
6. The system:
    a. Assigns the request a unique identification number (request ID);
    b. If the specific request:
        i. Involves an intra-faculty transaction only (e.g. asset from a faculty assigned to a user of that faculty) and if the user holds Role 2 from that faculty or Role 3; OR
        ii. Involves a faculty to faculty transaction (e.g. asset from a faculty assigned to a user from another faculty) and the user holds Role 3;
            1. The system sets the request status to *Approved*;
            2. The system issues the transaction implied by the request to the database;
        iii. Otherwise:
            1. The system sets the request status to *Pending*;
    c. Adds the request to the database including request ID, category (*Move asset to location*, *Move group to location*, *Assign equipment to user*, *Assign storage compartment to user* or *Assign group to user* as applicable), barcode, location, group ID, username, compartment number, status, originator username, faculty, department and level;
    d. Displays a message to the user with the request ID and status;
        i. The user selects the *Return* button;

**Post-conditions:**

1. The system displays the system page from which the *Requests* tab was accessed; and
2. The new request is stored in the database;

**Exceptions:**

1. The user cancels the request by selecting the *Cancel* button from the *Submit specific request* page:
    a. The system displays the original system page (from which the *Requests* tab was accessed); and

b. The system does not store any information associated with the cancelled request;
2. A database error is encountered:
   a. The system displays an error message.

*3.2.3.2.3 Search Request*

**Role:** All users;
**Pre-condition:** Authenticated session;
**Steps:**
1. The user selects the *Search request* button under the *Requests* tab (accessible from all system pages);
2. The system displays the *Search request* page;
3. The user provides the following (all optional) search information:
   a. Request ID;
   b. Category(s) – checkbox with choices *General administrative*, *General technical*, *Move asset to location*, *Move group to location*, *Assign equipment to user*, *Assign storage compartment to user* and *Assign group to user*;
   c. Description search string – short text;
   d. Status(s) – checkbox with choices *Pending, Closed* and *Approved*;
   e. Barcode;
   f. Location;
   g. Group ID;
   h. Username;
   i. Originator username;
   j. Originator faculty;
   k. Originator department;
   l. Originator level;
   m. Compartment number; and
   n. Closure note search string – short text;
4. The user selects the *Search* button;
5. The system builds the query based on the selected parameters and issues it to the database to perform:
   a. If the user has Role 0, the returned results are limited to those for which he is the originator;
   b. If the user has Role 1 and above, the returned results are limited to those for which Originator level is equal or lower than the user's role AND Originator department is the same as the user's;
6. The system displays the *Search results* page with the list of requests returned by the query with:
   a. Request ID;
   b. Category; and

c. Status;

**Post-condition:** The system displays the *Search results* page;

**Exceptions:**

1. The user cancels the search by selecting the *Return* button from the *Search request* page:
   a. The system displays the original system page (from which the *Requests* tab was accessed);
2. The user triggers a modified search by selecting the *Modify search* button from the *Search results* page:
   a. The system displays the *Search request* page with the last input value for all search information fields;
3. The user triggers a new search by selecting the *New search* button from the *Search results* page:
   a. The system displays the *Search request* page with all search information fields cleared;
4. The user clears all search information fields by selecting the *Clear search parameters* button from the *Search request* page:
   a. The system displays the *Search request* page with all search information fields cleared;
5. The query does not return any result:
   a. The system displays a "No results found" message;
6. A database error is encountered:
   a. The system displays an error message.

*3.2.3.2.4 View Request*

**Role:** All users;

**Pre-condition:** The system displays the *Search results* page with successful query results;

**Steps:**

1. The user selects a specific request using a radio button;
2. The user selects the *View request* button;
3. The system displays the *Request details* page with:
   a. Request ID;
   b. Category;
   c. Description;
   d. Status;
   e. Barcode;
   f. Location;
   g. Group ID;
   h. Username;
   i. Originator username;
   j. Originator faculty;

k. Originator department;
   l. Originator level;
   m. Compartment number; and
   n. Closure note;

**Post-condition:** The system displays the *Request details* page;

**Exceptions:**

1. The user returns to the *Search results* page by selecting the *Return* button from the *Request details* page:
   a. The system displays the *Search results* page with the results from the last query;
2. A database error is encountered:
   a. The system displays an error message.

*3.2.3.2.5 Close General Request*

**Role:** Designated user (one for each level based on the requester's level);

**Pre-condition:** The system displays the *Request details* page for a request of category *General administrative* or *General technical*;

**Steps:**

1. The user selects the *Close general request* button;
2. The system displays the *Close general request* page;
3. The user provides the following (mandatory) information:
   i. Closure note;
4. The user selects the *Submit* button;
5. The system, if no closure note has been provided, displays a message indicating the missing information and requesting the user to provide it:
   a. The user selects the *Return* button; and
   b. The system displays the *Close general request* page;
6. The system, if the user is the designated user:
   a. Sets the problem status to *Closed*;
   b. Updates the request database entry with the closure note and status change;
   c. Displays a message to the user with the request ID and status;
      i. The user selects the *Return* button;

**Post-conditions:**

1. The system displays the *Request details* page with updated information; and
2. The closed request is stored in the database;

**Exceptions:**

1. The user cancels the closure by selecting the *Cancel* button from the *Close general request* page:
   a. The system displays the *Request details* page; and
   b. The system does not store any updated information;
2. Upon submission, if the user is not the designated user, the system:

a. Does not store any updated information;
   b. Displays a message to the user stating he is not authorized to perform the operation;
      i. The user selects the *Return* button;
   c. The system displays the *Request details* page with previous information;
3. A database error is encountered:
   a. The system displays an error message.

*3.2.3.2.6 Approve Specific Request*

**Role:** Role 2 user with matching faculty or Role 3 user;
**Pre-condition:** The system displays the *Request details* page for a request of category *Move asset to location*, *Move group to location*, *Assign equipment to user*, *Assign storage compartment to user* or *Assign group to user*;
**Steps:**

1. The user selects the *Approve specific request* button;
2. If the specific request:
   a. Involves an intra-faculty transaction only (e.g. asset from a faculty assigned to a user of that faculty) and if the user holds Role 2 from that faculty or Role 3; OR
   b. Involves a faculty to faculty transaction (e.g. asset from a faculty assigned to a user from another faculty) and the user holds Role 3;
      i. The system sets the request status to *Approved*;
      ii. The system issues the transaction implied by the request to the database;
      iii. Updates the request database entry with status change; and
      iv. Displays a message to the user with the request ID and status;
         1. The user selects the *Return* button;

**Post-conditions:**

1. The system displays the *Request details* page with updated information; and
2. The closed request is stored in the database;

**Exceptions:**

1. Upon approval, if the specific request:
   a. Involves an intra-faculty transaction only (e.g. asset from a faculty assigned to a user of that faculty) and if the user <u>does not</u> hold Role 2 from that faculty or Role 3; OR
   b. Involves a faculty to faculty transaction (e.g. asset from a faculty assigned to a user from another faculty) and the user <u>does not</u> hold Role 3;
      i. The system does not store any updated information;
      ii. Displays a message to the user stating he is not authorized to perform the operation;
         1. The user selects the *Return* button;
      iii. The system displays the *Request details* page with previous information;

2. A database error is encountered:
    a. The system displays an error message.

### 3.2.3.3 Functional Requirements

*3.2.3.3.1 Submit General Request*

**[F.Req.1]** The system shall allow users to submit general requests.

**[F.Req.2]** The system shall assist a user, through structured dialogs, to describe the general request, including category from *General administrative* and *General technical*, and description.

**[F.Req.3]** The system shall, prior to proceeding with the submit request operation, display the problem related information including category and description and request a confirmation from the user.

**[F.Req.4]** The system shall, in the event where no description is provided, display a message to the user and withhold from performing the Submit general request operation.

**[F.Req.5]** The system shall allow a user to cancel a Submit general request operation.

**[F.Req.6]** The system shall, upon performing the Submit general request operation, create and store a request record with unique identification number, category, description, status set to *Pending*, originator name, faculty, department and level.

**[F.Req.7]** The system shall, upon successfully completing the Submit general request operation, display a message to the user including the request unique identification number.

*3.2.3.3.2 Submit Specific Request*

**[F.Req.8]** The system shall allow users to submit specific requests.

**[F.Req.9]** The system shall assist a user, through structured dialogs, to describe the specific request, including category (mandatory data) from *Move asset to location* (barcode and location), *Move group to location (*Group ID and location), *Assign equipment to user* (barcode and username), *Assign storage compartment to user* (barcode, compartment number and username), and *Assign group to user* (Group ID and username).

**[F.Req.10]** The system shall, if one or more mandatory field is missing, display a message to the user and withhold from performing the Submit specific request operation.

**[F.Req.11]** The system shall allow a user to cancel a Submit specific request operation.

**[F.Req.12]** The system shall, upon performing the Submit specific request operation, if the request approval criterion is met (involves an intra-faculty transaction only (e.g. asset from a faculty assigned to a user of that faculty) and if the user holds Role 2 from that faculty or Role 3; OR involves a faculty to faculty transaction (e.g. asset from a faculty assigned to a user from another faculty) and the user holds Role 3), issue the requested transaction to the database.

**[F.Req.13]** The system shall, upon performing the Submit specific request operation, create and store a request record with unique identification number, category, user provided data, status set to *Pending* or *Approved* as applicable, originator name, faculty, department and level.

**[F.Req.14]** The system shall, upon successfully completing the Submit specific request operation, display a message to the user including the request unique identification number and status.

*3.2.3.3.3 Search Request*

**[F.Req.15]** The system shall allow users to search requests.

**[F.Req.16]** The system shall assist a user, through structured dialogs, to specify search criteria, including request ID, category(s), description search string, status(s), barcode, location, group ID, username, originator username, faculty, department and level, compartment number and closure note search string.

**[F.Req.17]** The system shall allow a user to clear search criteria.

**[F.Req.18]** The system shall display the results of a request search including, for each returned result, request unique identification number, category, and status.

**[F.Req.19]** The system shall limit the returned results to:

a. If the user has Role 0, those for which he is the originator;
b. If the user has Role 1 and above, those for which Originator level is equal or lower than the user's role AND Originator department is the same as the user's.

**[F.Req.20]** The system shall, upon returning no results for a request search, display a message to the user.

**[F.Req.21]** The system shall allow a user to cancel a request search.

**[F.Req.22]** The system shall allow a user to modify the search criteria of a previously executed request search.

*3.2.3.3.4 View Request*

**[F.Req.23]** The system shall allow users to select a request for viewing.

**[F.Req.24]** The system shall display the request details including request ID, category, description, status, barcode, location, group ID, username, originator username, faculty, department and level, compartment number and closure note.

*3.2.3.3.5 Close General Request*

**[F.Req.25]** The system shall allow a designated user to close a general request.

**[F.Req.26]** The system shall assist a user, through structured dialogs, to provide a closure note.

**[F.Req.27]** The system shall, if a closure note is not provided, display a message to the user and withhold from performing the Close general request operation.

**[F.Req.28]** The system shall allow a user to cancel a Close general request operation.

**[F.Req.29]** The system shall, upon performing a Close general request operation, if the user is a designated user, update the request record including status set to *Closed* and closure note.

**[F.Req.30]** The system shall, upon successfully performing a Close general request operation, display a message to the user with request ID and status.

**[F.Req.31]** The system shall, upon attempting a Close general request operation, if the user is not a designated user, display a message to the user stating he is not authorized to perform the operation.

*3.2.3.3.6 Approve Specific Request*

**[F.Req.32]** The system shall allow a Role 2 user with faculty matching that of all the elements of a request and a Role 3 user to approve a specific request.

**[F.Req.33]** The system shall, upon performing the Approve specific request operation, if the request approval criterion is met (involves an intra-faculty transaction only (e.g. asset from a faculty assigned to a user of that faculty) and if the user holds Role 2 from that faculty or Role 3; OR involves a faculty to faculty transaction (e.g. asset from a faculty assigned to a user from another faculty) and the user holds Role 3), issue the requested transaction to the database.

**[F.Req.34]** The system shall, upon successfully performing an Approve specific request operation, update the request record including status set to *Approved*.

**[F.Req.35]** The system shall, upon successfully performing Approve specific request operation, display a message to the user with request ID and status.

**[F.Req.36]** The system shall, upon attempting a Close general request operation, if the request approval criterion is met, display a message to the user stating he is not authorized to perform the operation.

## 3.2.4 Physical Asset Management

The capabilities to manage (add, view, update, etc) physical assets is limited to Faculty users (Role 2) with *Manage physical assets* permission (for assets owned by their faculty only). University users (Role 3) with *Manage physical assets* permission have the capability to manage assets from all faculties.

Physical assets are defined as furniture (including storage unit) and equipment (including computer).

The physical assets management functionality allows the user to:

- Add, Search, View, Update (including location) physical assets inventory records;
- Assign a location to a physical asset;
- Assign an equipment to a user;
- Assign a storage compartment to a user;
- Create groups of physical assets
- View, update and delete groups of physical assets;
- Assign a location to a group; and
- Assign a group to a user.

### 3.2.4.1 Domain Model

The following figure illustrates the domain model applicable to the management of physical assets in the context of the IUfA UUIS.

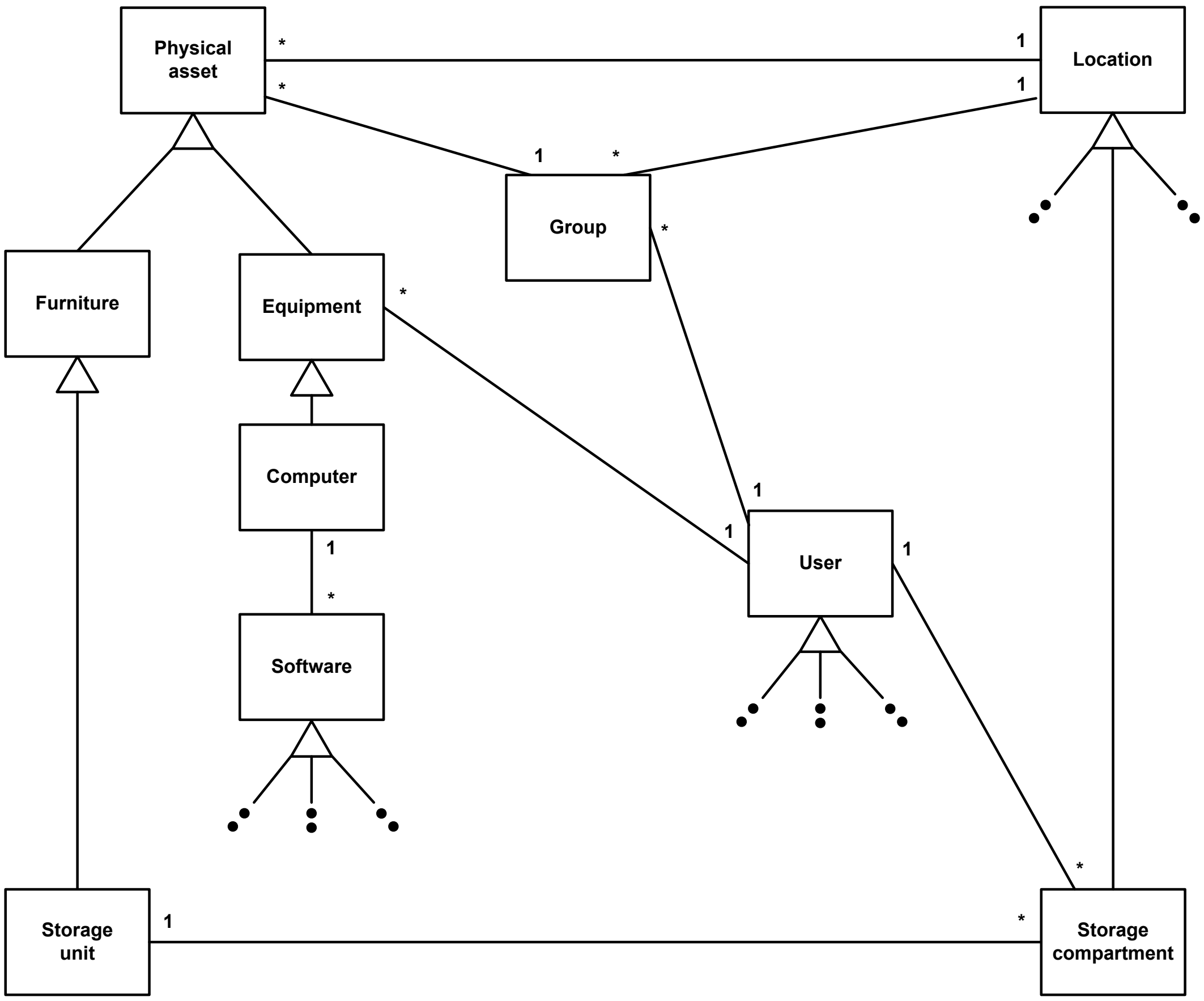

**Figure 8 -** Physical assets management – Domain model

### 3.2.4.2 Use Case Model

The following figure provides the Physical assets management use case diagram.

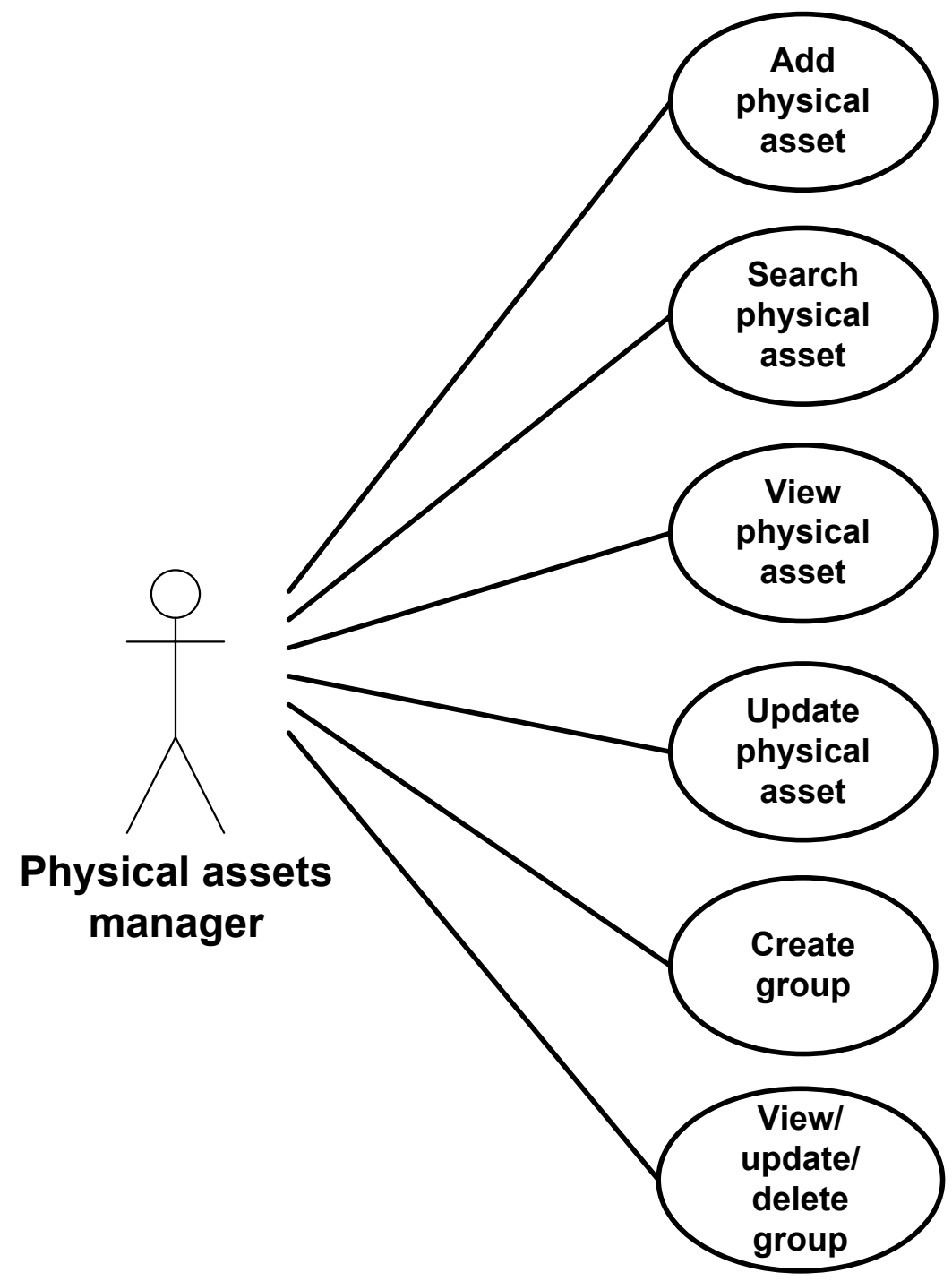

**Figure 9 -** Physical assets management – Use case diagram

The following paragraphs provide the detailed description for each use case.

*3.2.4.2.1 Add Physical Asset*

**Role:** Physical assets manager (Role 2), Physical assets manager (Role 3);
**Pre-conditions:**
1. Authenticated session; and
2. Faculty or University user with *Manage physical assets* permission;

**Steps:**
1. The physical assets manager selects the *Add asset* button under the *Physical assets* tab (accessible from all system pages);
2. The system displays the *Add asset* page;
3. The physical assets manager provides the following information:
    a. Barcode (mandatory) – the barcode is to be provided by the physical asset manager at the time of creation of the asset based on the stamp that he will affix to the physical asset;
    b. Owner (mandatory) – Faculty;

c. Purchase requisition number;
d. Purchase order number;
e. Legacy code;
f. Date purchased;
g. Warranty expiration;
h. Location;
i. Category (mandatory) – radio button with choices *Furniture* and *Equipment*;
j. Manufacturer;
k. Model;
l. Furniture description (if applicable):
    i. Furniture type (mandatory) from *Classroom front desk*, *Classroom desk*, *Classroom chair*, *Classroom desk/chair unit, Lab bench, Lab stool, Lab chair, Office Desk, Office chair, Office bookshelf, Table* and *Storage unit*;
    ii. Dimension;
    iii. Color; and
    iv. Finish;
    v. Additional fields specific to furniture type *Storage unit*:
        A. Storage unit type from *Locker*, *Filing cabinet* and *Shelving unit*;
        B. Number of compartments (mandatory), and:
            1. Compartment #1 user;
            2. Compartment #2 user;
            3. …
            4. Compartment #N user;
m. Equipment description (if applicable):
    i. Equipment type (mandatory) from *Computer*, *Screen*, *Keyboard*, *Mouse*, *Printer*, *Telephone*, *Photocopier*, *Projector*, *Screen*, *Amplifier*, *Microphone* and *Speaker*;
    ii. Serial number;
    iii. Equipment assigned user;
    iv. Additional fields specific to equipment type *Computer*:
        A. Computer type from *Tower*, *Laptop* and *Server*;
        B. Processor;
        C. MAC address;
        D. Hard drive capacity;
        E. Non-volatile memory; and
        F. Volatile memory;

4. The physical assets manager selects the *Submit* button;
5. The system:
    a. If one or more mandatory field is missing displays a message indicating the missing information and requesting the physical assets manager to provide it; and/or

b. In the case where the Physical Asset Manager is a Faculty user, if the specified Faculty differs from the Faculty in his profile, displays a message indicating the discrepancy;
      i. The physical assets manager selects the *Return* button; and
      ii. The system displays the *Add asset* page with current information;
6. The system displays a message with the information provided by the physical assets manager requesting confirmation:
   a. The physical assets manager selects the *Confirm* button; or
   b. The physical assets manager selects the *Return* button;
      i. The system displays the *Add asset* page with current information;
7. The system:
   a. Assigns the physical asset a unique identification number (asset ID);
   b. Sets the physical asset status to *In-stock*;
   c. Adds the physical asset to the database including asset ID, barcode, owner, legacy code, date purchased, warranty expiration, manufacturer, model, category and all furniture or equipment description fields, whichever is applicable; and
   d. Displays a message to the physical assets manager with the asset ID;
      i. The physical assets manager selects the *Return* button;

**Post-conditions:**

1. The system displays the *Add asset* page with all fields cleared; and
2. The new asset record is stored in the database;

**Exceptions:**

1. The physical assets manager cancels the physical asset addition by selecting the *Cancel* button from the *Add asset* page:
   a. The system displays the original system page (from which the *Physical assets* tab was accessed); and
   b. The system does not store any information associated with the physical asset addition;
2. A database error is encountered:
   a. The system displays an error message.

*3.2.4.2.2 Search Physical Asset*

**Role:** Physical assets manager (Role 2), Physical assets manager (Role 3);

**Pre-conditions:**

1. Authenticated session; and
2. Faculty or University user with *Manage physical assets* permission;

**Steps:**

1. The physical assets manager selects the *Search asset* button under the *Physical assets* tab (accessible from all system pages);
2. The system displays the *Search asset* page;

3. The physical assets manager provides the following (all optional) search information:
   a. Asset ID;
   b. Barcode;
   c. Owner – Faculty;
   d. Purchase requisition number;
   e. Purchase order number;
   f. Legacy code;
   g. Date purchased date range;
   h. Warranty expiration date range;
   i. Location;
   j. Group ID;
   k. Manufacturer;
   l. Model;
   m. Category – checkbox with choices *Furniture* and *Equipment*;
   n. Furniture type(s) – checkbox with choices *Classroom front desk*, *Classroom desk*, *Classroom chair*, *Classroom desk/chair unit, Lab bench, Lab stool, Lab chair, Office Desk, Office chair, Office bookshelf, Table* and *Storage unit*;
   o. Storage unit type(s) – checkbox with choices *Locker*, *Filing cabinet* and *Shelving unit*;
   p. Equipment type(s) – checkbox with choices *Computer*, *Screen*, *Keyboard*, *Mouse*, *Printer*, *Telephone*, *Photocopier*, *Projector*, *Screen*, *Amplifier*, *Microphone* and *Speaker*;
   q. Equipment serial number;
   r. Equipment assigned user;
   s. Computer type(s) – checkbox with choices *Tower*, *Laptop* and *Server*;
   t. MAC address;
   u. Hard drive capacity range in Gbytes;
   v. Non-volatile memory range in Mbytes; and
   w. Volatile memory range in Mbytes;
   x. Status(s) – checkbox with choices *In-stock, Assigned, Damaged, In-repair, Lost* and *Retired*;
4. The physical assets manager selects the *Search* button;
5. The system builds the query based on the selected parameters and issues it to the database to perform. In the case the Physical Asset Manager is a Faculty user, the search is limited to the Faculty in his profile, regardless of the above choice;
6. The system displays the *Search results* page with the list of physical asset records returned by the query with:
   a. Asset ID;
   b. Barcode;
   c. Category;
   d. Location;
   e. Manufacturer;

f. Model;
g. Furniture type (if applicable);
h. Equipment type (if applicable);
i. Equipment assigned user; and
j. Status;

**Post-condition:** The system displays the *Search results* page;
**Exceptions:**

1. The physical assets manager cancels the search by selecting the *Return* button from the *Search asset* page:
   a. The system displays the original system page (from which the *Physical assets* tab was accessed);
2. The physical assets manager triggers a modified search by selecting the *Modify search* button from the *Search results* page:
   a. The system displays the *Search asset* page with the last input value for all search information fields;
3. The physical assets manager triggers a new search by selecting the *New search* button from the *Search results* page:
   a. The system displays the *Search asset* page with all search information fields cleared;
4. The physical assets manager clears all search information fields by selecting the *Clear search parameters* button from the *Search asset* page:
   a. The system displays the *Search asset* page with all search information fields cleared;
5. The query does not return any result:
   a. The system displays a "No results found" message;
6. A database error is encountered:
   a. The system displays an error message.

*3.2.4.2.3 View Physical Asset*

**Role:** Physical assets manager (Role 2), Physical assets manager (Role 3);
**Pre-condition:** The system displays the *Search results* page with successful query results;
**Steps:**

1. The physical assets manager selects a specific physical asset using a radio button;
2. The physical assets manager selects the *View asset* button;
3. The system displays the *Asset details* page with:
   a. Asset ID;
   b. Barcode;
   c. Owner – Faculty;
   d. Purchase requisition number;
   e. Purchase order number;
   f. Legacy code if applicable;

g. Date purchased;
h. Warranty expiration;
i. Location;
j. Group ID;
k. Manufacturer;
l. Model;
m. Category;
n. Furniture description (if applicable):
    i. Furniture type;
    ii. Dimension;
    iii. Color; and
    iv. Finish;
    v. Additional fields specific to furniture type *Storage unit* (if applicable):
        1. Storage unit type;
        2. Number of compartments, and:
            a. Compartment #1 user;
            b. Compartment #2 user;
            c. …
            d. Compartment #N user;
o. Equipment description (if applicable):
    i. Equipment type;
    ii. Serial number;
    iii. Equipment assigned user;
    iv. Additional fields specific to equipment type *Computer* (if applicable):
        1. Computer type;
        2. Processor;
        3. MAC address;
        4. Hard drive capacity;
        5. Non-volatile memory; and
        6. Volatile memory;
p. Status;

**Post-condition:** The system displays the *Asset details* page;
**Exceptions:**

1. The physical assets manager returns to the *Search results* page by selecting the *Return* button from the *Asset details* page:
    a. The system displays the *Search results* page with the results from the last query;
2. A database error is encountered:
    a. The system displays an error message.

*3.2.4.2.4 Update Physical Asset*

**Role:** Physical assets manager (Role 2), Physical assets manager (Role 3);
**Pre-condition:** The system displays the *Asset details* page;

**Steps:**

1. The physical assets manager selects the *Update physical asset* button;
2. The system displays the *Edit physical asset* page with:
    a. Un-editable fields:
        i. Asset ID;
        ii. Barcode;
        iii. Purchase requisition number;
        iv. Purchase order number;
        v. Manufacturer;
        vi. Model;
        vii. Category;
        viii. Furniture type if applicable;
        ix. Storage unit type if applicable;
        x. Equipment type if applicable;
        xi. Equipment serial number if applicable;
        xii. Computer type if applicable;
    b. Editable fields:
        i. Legacy code;
        ii. Owner – Faculty; Note: Editing of this field is limited to Physical Asset Manager (Role 3);
        iii. Date purchased;
        iv. Warranty expiration;
        v. Location;
        vi. Status from *In-stock, Assigned, Damaged, In-repair, Lost* and *Retired*;
        vii. Furniture dimension if applicable;
        viii. Furniture color if applicable;
        ix. Furniture finish if applicable;
        x. If applicable:
            1. Compartment #1 user;
            2. Compartment #2 user;
            3. …
            4. Compartment #N user;
        xi. Equipment assigned user;
        xii. Computer processor if applicable;
        xiii. Computer MAC address if applicable;
        xiv. Computer hard drive capacity if applicable;
        xv. Computer non-volatile memory if applicable; and
        xvi. Computer volatile memory if applicable;
3. The physical assets manager edits/updates the above fields as applicable;
4. The physical assets manager selects the *Submit* button;
5. The system displays a message with the information provided by the physical assets manager requesting confirmation:
    a. The physical assets manager selects the *Confirm* button; or

b. The physical assets manager selects the *Return* button;
      i. The system displays the *Update physical asset* page with current information;
6. The system:
   a. Updates the physical asset database entry with the applicable changes;

**Post-conditions:**

1. The system displays the *Asset details* page with updated information; and
2. The updated physical asset record is stored in the database;

**Exceptions:**

1. The user cancels the update by selecting the *Cancel* button from the *Edit physical asset* page:
   a. The system displays the *Asset details* page; and
   b. The system does not store any updated information;
2. A database error is encountered:
   a. The system displays an error message.

*3.2.4.2.5 Create Group*

**Role:** Physical assets manager (Role 2), Physical assets manager (Role 3);

**Pre-conditions:**

1. Authenticated session; and
2. Faculty or University user with *Manage physical assets* permission;

**Steps:**

1. The physical assets manager selects the *Create group* button under the *Physical assets* tab (accessible from all system pages);
2. The system displays the *Create group* page;
3. The physical assets manager provides the following information:
   a. Group assets (one mandatory):
      i. Asset #1 – Asset ID;
      ii. Asset #2 – Asset ID;
      iii. …
      iv. Asset #N – Asset ID;
   b. Group location;
   c. Group assigned user;
4. The physical assets manager selects the *Submit* button;
5. The system:
   a. If the submitted group does not contain at least one asset, displays a message and requests the physical assets manager to provide a minimum of one asset; and/or
   b. In the case where the Physical Asset Manager is a Faculty user, if one or more of the specified asset belongs to a different Faculty than the Faculty in his profile, displays a message indicating the discrepancy;

i. The physical assets manager selects the *Return* button; and
   ii. The system displays the *Create group* page with current information;
6. The system displays a message with the information provided by the physical assets manager requesting confirmation:
   a. The physical assets manager selects the *Confirm* button; or
   b. The physical assets manager selects the *Return* button;
      i. The system displays the *Create group* page with current information;
7. The system:
   a. Assigns the group a unique identification number (group ID);
   b. Adds the group to the database including group ID, asset ID(s), location if provided and assigned user if provided;
   c. Updates the record for every individual asset of the group with, if provided, the group location;
   d. Updates the record for every individual asset of the group with, if provided, the group assigned user;
   e. Displays a message to the physical assets manager with the group ID;
      i. The physical assets manager selects the *Return* button;

**Post-conditions:**

1. The system displays the *Create group* page with all fields cleared;
2. The new group record is stored in the database; and
3. The updated physical asset records are stored in the database;

**Exceptions:**

1. The physical assets manager cancels the group creation by selecting the *Cancel* button from the *Create group* page:
   a. The system displays the original system page (from which the *Physical asset* tab was accessed); and
   b. The system does not store any information associated with the group creation;
2. A database error is encountered:
   a. The system displays an error message.

*3.2.4.2.6 View/Update/Delete Group*

**Role:** Physical assets manager (Role 2), Physical assets manager (Role 3);

**Pre-conditions:**

1. Authenticated session; and
2. Faculty or University user with *Manage physical assets* permission;

**Steps:**

1. The physical assets manager selects the *View/update/delete group* button under the *Physical assets* tab (accessible from all system pages);
2. The system displays the *Retrieve group* page;
3. The physical assets manager provides the group ID;
4. The physical assets manager selects the *Retrieve* button;

5. The system builds the query and issues to the database to perform;
6. The system displays the *Group details* page with (all editable):
    a. Group assets:
        i. Asset #1 – Asset ID;
        ii. Asset #2 – Asset ID;
        iii. …
        iv. Asset #N – Asset ID;
    b. Group location; and
    c. Group assigned user;
7. The physical assets manager edits/updates the above fields as applicable;
8. The physical assets manager selects the *Submit* button;
9. The system, in the case where the Physical Asset Manager is a Faculty user, if one or more of the specified asset belongs to a different Faculty than the Faculty in his profile, displays a message indicating the discrepancy;
    i. The physical assets manager selects the *Return* button; and
    ii. The system displays the *Group details* page with current information;
10. The system displays a message with the information provided by the physical assets manager requesting confirmation:
    a. The physical assets manager selects the *Confirm* button; or
    b. The physical assets manager selects the *Return* button;
        i. The system displays the *Group details* page with current information;
11. If the updated group contains at least one asset, the system:
    a. Updates the group database entry with the applicable change(s);
    b. Updates the record for every individual asset of the group with, if updated, the group location;
    c. Updates the record for every individual asset of the group with, if updated, the group assigned user;
    d. Displays a message to the physical assets manager with the group ID;
        i. The physical assets manager selects the *Return* button;
12. If the updated group does not contain at least one asset, the system:
    a. Deletes the group database entry;
    b. Leaves intact (location and assigned user) the record for every individual asset previously of the group;
    c. Displays a message to the physical assets manager with the group ID;
        i. The physical assets manager selects the *Return* button;

**Post-conditions:**

1. The system displays the *Retrieve group* page;
2. The updated group record and the updated physical asset records are stored in the database; or
3. The group database entry is deleted and the physical asset records are left intact;

**Exceptions:**

1. The physical asset manager cancel the retrieval by selecting the *Cancel* button from the *Retrieve group* page;

a. The system displays the original system page (from which the *Physical assets* tab was accessed);
2. The physical assets manager returns to the *Retrieve group* page by selecting the *Return* button from the *Group details* page:
    a. The system displays the *Retrieve group* page;
3. The query does not return any result:
    a. The system displays a "No results found message";
    b. The physical assets manager selects the *Return* button;
    c. The system displays the *Retrieve group* page;
4. A database error is encountered:
    a. The system displays an error message.

### 3.2.4.3 Functional Requirements

*3.2.4.3.1 Add Physical Asset*

**[F.Req.1]** The system shall allow Faculty users with *Manage physical assets* permission to add physical assets to the inventory;

**[F.Req.2]** The system shall assist a user, through structured dialogs, to describe the added physical asset, including barcode, owner (faculty), purchase requisition number, purchase order number, legacy code, date purchased, warranty expiration, location, manufacturer, model, category from *furniture* and *equipment*, furniture description if applicable (furniture type, dimension, color and finish), storage unit description if applicable (type, number of compartments and assigned users), equipment description if applicable (equipment type, serial number and equipment assigned user) and computer description if applicable (computer type, processor, MAC address, hard drive capacity, non-volatile memory and volatile memory).

**[F.Req.3]** The system shall, prior to proceeding with the *add physical asset* operation, display the physical asset related information and request a confirmation from the user.

**[F.Req.4]** The system shall, in the event where no barcode, owner, category, furniture type (if applicable), number of compartments (if applicable) and/or equipment type (if applicable) are provided, display a message to the user and withhold from performing the *add physical asset* operation.

**[F.Req.5]** The system shall, in the case where the Physical Asset Manager is a Faculty user, if the specified Faculty differs from the Faculty in his profile, display a message to the user and withhold from performing the *add physical asset* operation.

**[F.Req.6]** The system shall allow a user to cancel an *add physical asset* operation.

**[F.Req.7]** The system shall, upon performing the *add physical asset* operation, create and store a physical asset record with unique ID number, status reflecting its added state and user provided information.

**[F.Req.8]** The system shall, upon successfully completing the *add physical asset* operation, display a message to the user including the physical asset record unique ID number.

*3.2.4.3.2 Search Physical Asset*

**[F.Req.9]** The system shall allow Faculty users with *Manage physical assets* permission to search physical assets.

**[F.Req.10]** The system shall assist a user, through structured dialogs, to specify search criteria, including physical asset unique ID number, barcode, owner(s), purchase requisition number, purchase order number, legacy code, date purchased date range, warranty expiration date range, location,

manufacturer, model, category, furniture type(s), storage unit type(s), equipment type(s), equipment serial number, equipment assigned user, computer type(s), MAC address, hard drive capacity range in Gbytes, non-volatile memory range in Mbytes, volatile memory range in Mbytes and status(s).

**[F.Req.11]** The system shall allow a user to clear search criteria.

**[F.Req.12]** The system shall, in the case where the Physical Asset Manager is a Faculty user, limit the search results to the Faculty in his profile.

**[F.Req.13]** The system shall display the results of a physical asset search including, for each returned result, physical asset unique ID number, barcode, manufacturer, model, category, location, furniture type (if applicable), equipment type (if applicable), equipment assigned user and status.

**[F.Req.14]** The system shall, upon returning no results for a physical asset search, display a message to the user.

**[F.Req.15]** The system shall allow a user to cancel a physical asset search.

**[F.Req.16]** The system shall allow a user to modify the search criteria of a previously executed physical asset search.

*3.2.4.3.3 View Physical Asset*

**[F.Req.17]** The system shall allow Faculty users with *Manage physical assets* permission to select a physical asset for viewing.

**[F.Req.18]** The system shall display the physical asset details including physical asset unique ID number, barcode, owner, purchase requisition number, purchase order number, legacy code, date purchased, warranty expiration, location, manufacturer, model, category, furniture description if applicable (furniture type, dimension, color and finish), storage unit description if applicable (type, number of compartments and assigned users), equipment description if applicable (equipment type, serial number and equipment assigned user), computer description if applicable (computer type, processor, MAC address, hard drive capacity, non-volatile memory and volatile memory) and status.

*3.2.4.3.4 Update Physical Asset*

**[F.Req.19]** The system shall allow Faculty users with *Manage physical assets* permission to update a physical asset record.

**[F.Req.20]** The system shall allow the user to update a physical asset record including legacy code, owner, date purchased, warranty expiration, location, status from *In-stock, Assigned, Damaged, In-repair, Lost* and *Retired*, furniture dimension (if applicable), furniture color (if applicable), furniture finish (if applicable), compartment #1 to compartment #N assigned user (if applicable), equipment assigned user (if applicable), computer processor (if

applicable), computer MAC address (if applicable), computer hard drive capacity (if applicable), computer non-volatile memory (if applicable) and computer volatile memory (if applicable).

**[F.Req.21]** The system shall limit the editing of the owner field to user with Physical Asset Manager (Role 3);

**[F.Req.22]** The system shall, prior to performing the *update physical asset record* operation, display the update information and request a confirmation from the user.

**[F.Req.23]** The system shall allow a user to cancel an *update physical asset record* operation.

**[F.Req.24]** The system shall, following confirmation from the user, modify the physical asset record to include user provided update information.

*3.2.4.3.5 Create Group*

**[F.Req.25]** The system shall allow Faculty users with *Manage physical assets* permission to group one or more physical assets under a unique group identifier.

**[F.Req.26]** The system shall allow a user to specify the location and assigned user of a group.

**[F.Req.27]** The system shall, in the event where no asset is provided, display a message to the user and withhold from performing the *create group* operation.

**[F.Req.28]** The system shall, in the case where the Physical Asset Manager is a Faculty user, if one or more of the specified asset belongs to a different Faculty than the Faculty in his profile, display a message to the user and withhold from performing the *create group* operation.

**[F.Req.29]** The system shall, prior to performing the *create group* operation, display the group information and request a confirmation from the user.

**[F.Req.30]** The system shall allow a user to cancel a *create group* operation.

**[F.Req.31]** The system shall, upon performing the *create group* operation, create and store a group record with unique ID number, asset ID number(s), location and assigned user.

**[F.Req.32]** The system shall, upon performing the *create group* operation, update the record for every individual asset in the group with, if provided, the group location and assigned user.

**[F.Req.33]** The system shall, upon successfully completing the *create group* operation, display a message to the user including the group unique ID number.

*3.2.4.3.6 View/Update/Delete Group*

**[F.Req.34]** The system shall allow Faculty users with *Manage physical assets* permission to view, update and delete group records.

**[F.Req.35]** The system shall allow the user to update a physical asset record including asset ID(s), location and assigned user.

**[F.Req.36]** The system shall, in the case where the Physical Asset Manager is a Faculty user, if one or more of the specified asset belongs to a different Faculty than the Faculty in his profile, display a message to the user and withhold from performing the group update operation.

**[F.Req.37]** The system shall, prior to performing the group update operation, display the update information and request a confirmation from the user.

**[F.Req.38]** The system shall allow a user to cancel a group update operation.

**[F.Req.39]** The system shall, upon update of a group (i.e. update information contains at least one asset), modify the group record to include user provided update information and update the record for every individual asset in the group with, if provided, the group location and assigned user.

**[F.Req.40]** The system shall, upon deletion of a group (i.e. update information does not contain at least one asset), delete the group database entry and leave intact (location and assigned user) the record for every individual asset previously of the group.

**[F.Req.41]** The system shall, upon successfully completing the group update operation, display a message to the user including the group unique ID number.

### 3.2.5 Software Management

The capability to manage software is limited to University users (Role 3) with *Manage software* permission, a role/permission attributed to IT department employees.

The acquisition of software is initiated by the submission, by a faculty member, of a general technical request for the installation of software (refer to the Requests section of the SRS). General requests are monitored by designated individuals from the IT department who will take the appropriate actions.

The software management functionality allows the user to:

- Add software to the inventory following the acquisition of new software and licenses;
- Perform searches based on Software and License information;
- Add/Discontinue software licenses; and
- Assign/Un-assign licenses to/from computer(s).

The software management functionality also includes a system function that provides automated notification of license near expiry.

### 3.2.5.1 Use Case Model

The following figure provides the Software management use case diagram.

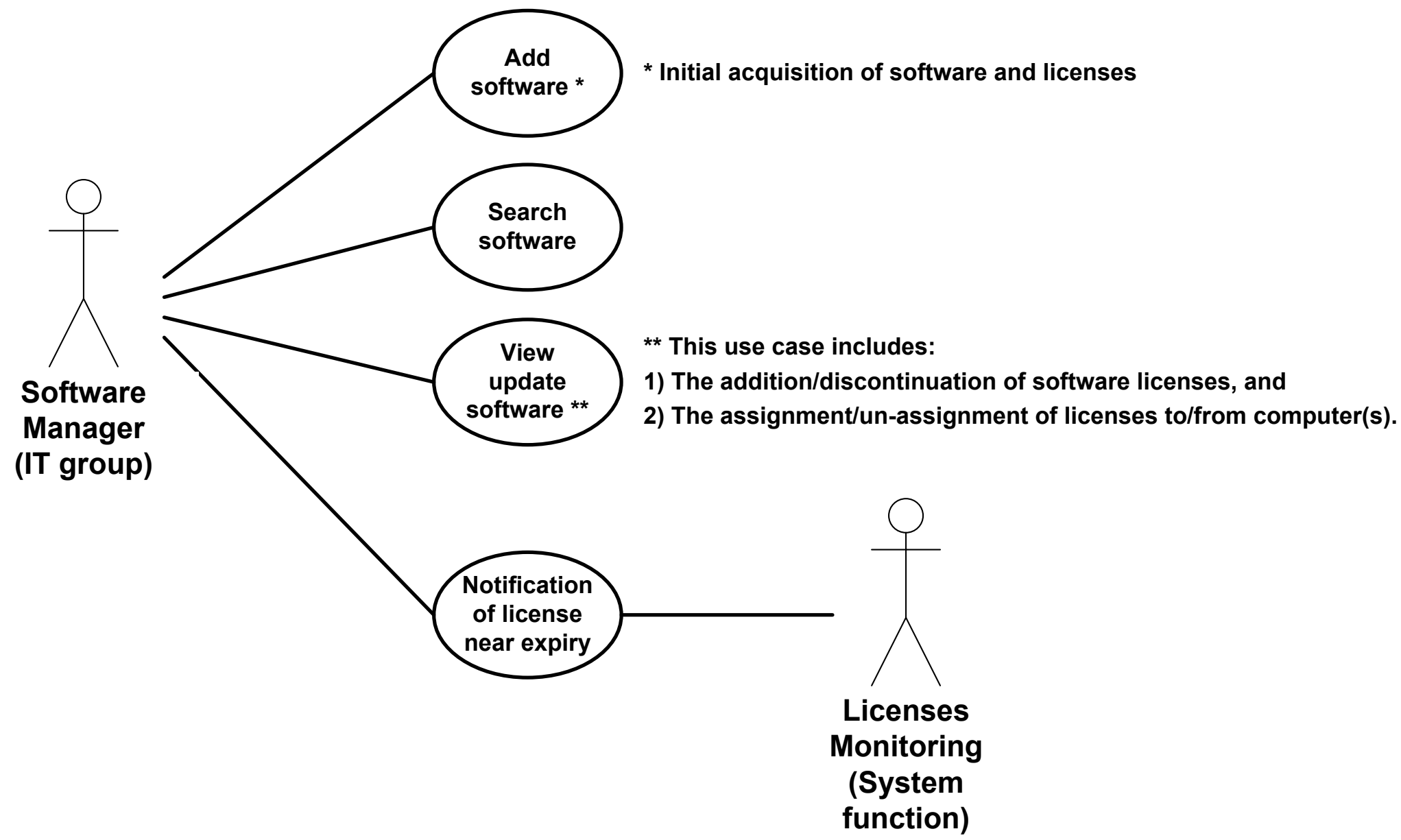

**Figure 10 -** Software management – Use case diagram

The following paragraphs provide the detailed description for each use case.

*3.2.5.1.1 Add Software (Acquisition of New Software and Licenses)*

**Role:** Software manager (Role 3)
**Pre-conditions:**
1. Authenticated session; and
2. IT department designated users (Role 3) with *Manage software* permission;

**Steps:**
1. The software manager selects the *Add software* button under the *Software* tab (accessible from all system pages);
2. The system displays the *Add software* page;
3. The software manager provides the following information:
    a. Name (mandatory);
    b. Vendor – drop-down menu;
    c. Category – drop-down menu;
    d. Description – text field;
    e. License #1:
        i. Version (mandatory);
        ii. Serial number (mandatory);

iii. License type (mandatory) – drop-down;
iv. Maximum number of installations (mandatory);
v. Active/inactive (mandatory) – Boolean (default = active);
vi. Purpose – drop-down;
vii. Requirements – text field;
viii. Requisition number;
ix. Purchase order number; and
x. Expiry date;

f. License #2:
   i. Version (mandatory);
   ii. …
   x. Expiry date;
g. License #N:
   i. Version (mandatory);
   ii. …
   x. Expiry date;

4. The software manager selects the *Submit* button;
5. The system, if one or more mandatory field is missing, displays a message indicating the missing information and requesting the software manager to provide it;
   a. The software manager selects the *Return* button; and
   b. The system displays the *Add software* page with current information;
6. The system displays a message with the information provided by the software manager requesting confirmation:
   a. The software manager selects the *Confirm* button; or
   b. The software manager selects the *Return* button;
      i. The system displays the *Add software* page with current information;
7. The system:
   a. Assigns the new software a unique identification number (software ID);
   b. Adds the new software to the database including software ID, name, vendor, category and description;
   c. Assigns each new licence a unique identification number (license ID);
   d. Adds each new license to the database including license ID, version, serial number, license type, maximum number of installations, purpose, requirements, requisition number, Purchase order number and expiry date; and
   e. Displays a message to the physical assets manager with the software ID and each license ID;
      i. The physical assets manager selects the *Return* button;

**Post-conditions:**

2. The system displays the *Add software* page with all fields cleared; and
3. The new software record and license records are stored in the database;

**Exceptions:**

1. The software manager cancels the software addition by selecting the *Cancel* button from the *Add software* page:
   a. The system displays the original system page (from which the *Software* tab was accessed); and
   b. The system does not store any information associated with the software addition;
2. A database error is encountered:
   a. The system displays an error message.

*3.2.5.1.2 Search Software*

**Role:** Software manager (Role 3)
**Pre-conditions:**
1. Authenticated session; and
2. IT department designated users (Role 3) with *Manage software* permission;

**Steps:**
1. The software manager selects the *Add software* button under the *Software* tab (accessible from all system pages);
2. The system displays the *Add software* page;
3. The software manager provides the following information:
   a. Name (mandatory);
   b. Vendor – drop-down menu;
   c. Category – drop-down menu;
   d. Description – text field;
   e. License #1:
      i. Version (mandatory);
      ii. Serial number (mandatory);
      iii. License type (mandatory) – drop-down;
      iv. Maximum number of installations (mandatory);
      v. Active/inactive (mandatory) – Boolean (default = active);
      vi. Purpose – drop-down;
      vii. Requirements – text field;
      viii. Requisition number;
      ix. Purchase order number; and
      x. Expiry date;
   f. License #2:
      i. Version (mandatory);
      ii. …
      xi. Expiry date;
   g. License #N:
      i. Version (mandatory);
      ii. …

xi. Expiry date;
4. The software manager selects the *Submit* button;
5. The system, if one or more mandatory field is missing, displays a message indicating the missing information and requesting the software manager to provide it;
    a. The software manager selects the *Return* button; and
    b. The system displays the *Add software* page with current information;
6. The system displays a message with the information provided by the software manager requesting confirmation:
    a. The software manager selects the *Confirm* button; or
    b. The software manager selects the *Return* button;
        i. The system displays the *Add software* page with current information;
7. The system:
    a. Assigns the new software a unique identification number (software ID);
    b. Adds the new software to the database including software ID, name, vendor, category and description;
    c. Assigns each new licence a unique identification number (license ID);
    d. Adds each new license to the database including license ID, version, serial number, license type, maximum number of installations, purpose, requirements, requisition number, Purchase order number and expiry date; and
    e. Displays a message to the physical assets manager with the software ID and each license ID;
        i. The physical assets manager selects the *Return* button;

**Post-conditions:**

1. The system displays the *Add software* page with all fields cleared; and
2. The new software record and license records are stored in the database;

**Exceptions:**

1. The software manager cancels the software addition by selecting the *Cancel* button from the *Add software* page:
    a. The system displays the original system page (from which the *Software* tab was accessed); and
    b. The system does not store any information associated with the software addition;
2. A database error is encountered:
    a. The system displays an error message.

*3.2.5.1.3 View/Update Software*

**Note:** This use case includes:

1) The addition/discontinuation of software licenses, and

2) The assignment/un-assignment of licenses to/from computer(s).

**Role:** Software manager (Role 3);
**Pre-condition:** The system displays the *Search results* page with successful query results;
**Steps:**

1. The software manager selects a specific software item using a radio button;
2. The physical assets manager selects the *View/update software* button;
3. The system displays the *Software details* page with:
    a. Name;
    b. Vendor (editable);
    c. Category (editable);
    d. Description (editable);
    e. License #1:
        i. Version;
        ii. Serial number;
        iii. License type (editable);
        iv. Maximum number of installations;
        v. Active/inactive (editable);
        vi. Purpose (editable);
        vii. Requirements (editable);
        viii. Requisition number (editable);
        ix. Purchase order number (editable); and
        x. Expiry date (editable);
        xi. Installed on:
            A. Asset ID #1 (editable);
            B. Asset ID #2 (editable);
            C. …
            D. Asset ID #maxInstalls (editable);
    f. License #2:
        i. Version;
        ii. …
        ix. Expiry date (editable);
        x. Installed on:
            A. Asset ID #1 (editable);
            B. Asset ID #2 (editable);
            C. …
            D. Asset ID #maxInstalls (editable);
    g. License #N:
        i. Version (mandatory);
        ii. …
        ix. Expiry date (editable);
        x. Installed on:
            A. Asset ID #1 (editable);

B. Asset ID #2 (editable);
C. …
D. Asset ID #maxInstalls (editable);

4. The software manager edits/updates the above fields as applicable:
   a. The page provides the user with the capability to add license(s);
   b. The page provides the user with the capability to set the maximum number of installations for a license upon addition only (not on update); and
   c. A license is discontinued by setting the Active/inactive parameter to *Inactive*;
5. The software manager selects the *Submit* button;
6. The system displays a message with the information provided by the software manager requesting confirmation:
   a. The software manager selects the *Confirm* button; or
   b. The software manager selects the *Return* button;
      i. The system displays the *Software details* page with current information;
7. The system:
   a. Updates the software and license(s) database entry(s) with the applicable changes;

**Post-conditions:**

1. The system displays the *Software details* page with updated information; and
2. The updated software record is stored in the database;

**Exceptions:**

1. The software manager returns to the *Search results* page by selecting the *Return* button from the *Software details* page:
   a. The system displays the *Search results* page with the results from the last query;
2. A database error is encountered:
   a. The system displays an error message.

*3.2.5.1.4 Notification of license Near Expiry*

**Roles:** Software manager (Role 3); License monitoring – System function;
**Pre-condition:** None;
**Steps:**

1. The software manager selects the *Enable notification of license near expiry* button under the *Software* tab (accessible from all system pages);
2. The system displays the *Notification of license near expiry* page;
3. The software manager provides the following information:
   a. Enable/disable notification of license near expiry – Boolean (default = disabled); and
   b. Set threshold in days (ex: expiry within 30 days);
4. The software manager selects the *Set* button;
5. The system:

a. Enables, if applicable, the notification of license near expiry function.
6. The notification of license near expiry function:
    a. Periodically (daily) monitor the expiry date of all active software licences; and
    b. Issues a notification to the software manager with the list of software licenses (serial #) for which the expiry date falls within the set threshold (in days) of the current date, if any exists;

**Post-conditions:**

1. The system displays the original system page (from which the *Software* tab was accessed); and
2. The notification of license near expiry function is enabled or disabled, as applicable.

**Exceptions:**

1. The user cancels the update by selecting the *Cancel* button from the *Notification of license near expiry*:
    a. The system displays the original system page (from which the *Software* tab was accessed); and
    b. The notification of license near expiry function enabled/disabled state remains unchanged;
2. A database error is encountered:
    a. The system displays an error message.

### 3.2.5.2 Functional Requirements

*3.2.5.2.1 Add Software*

**[F.Req.1]** The system shall show a list of software categories

**[F.Req.2]** The system shall allow the user to manually add a software title into the software collection database

*3.2.5.2.2 Search Software*

**[F.Req.3]** The system shall provide the user with a basic search option to look for software

**[F.Req.4]** The system shall provide the user with an advanced search option to look for software

*3.2.5.2.3 View/Update Software*

**[F.Req.5]** The system shall show the list of software available for the university community

**[F.Req.6]** The system shall allow the user to update data related to a software

*3.2.5.2.4 Notification of License Near Expiry*

**[F.Req.7]** The system shall notify a user that a software license is about to expire

**[F.Req.8]** The system shall send an e-mail warning that a software license is about to expire

## 3.2.6 Manage Locations

The purpose of this module is to manage the different locations that constitute a university. Different location types are Building, lab, classroom, office, etc

### 3.2.6.1 Domain Model

The following figure illustrates the Domain Model applicable to Manage Locations in the context of the IUfA UUIS.

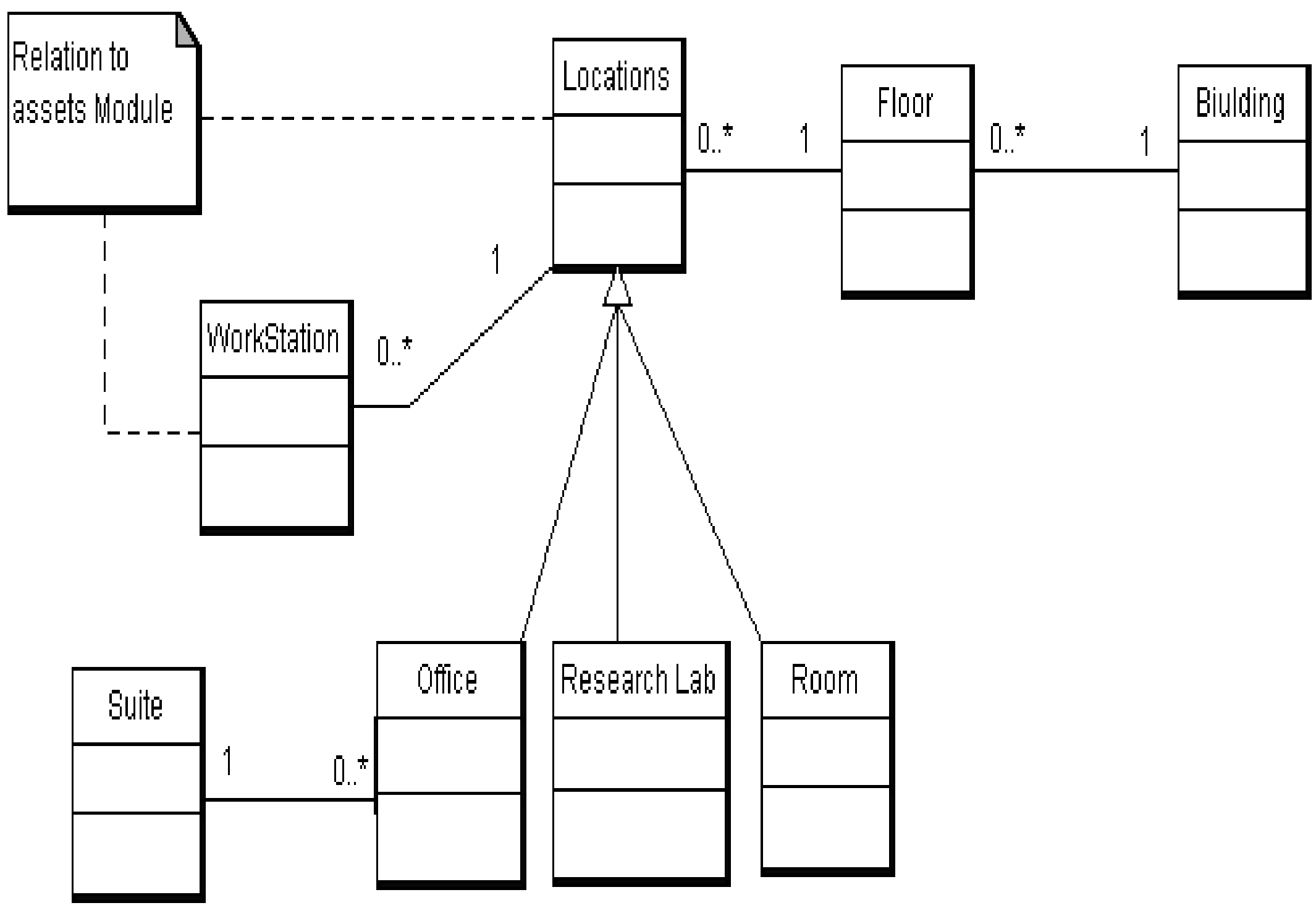

**Figure 11 –** Manage Locations – Domain model

### 3.2.6.2 Use Case Model

The following figure provides the Manage Location use case diagram.

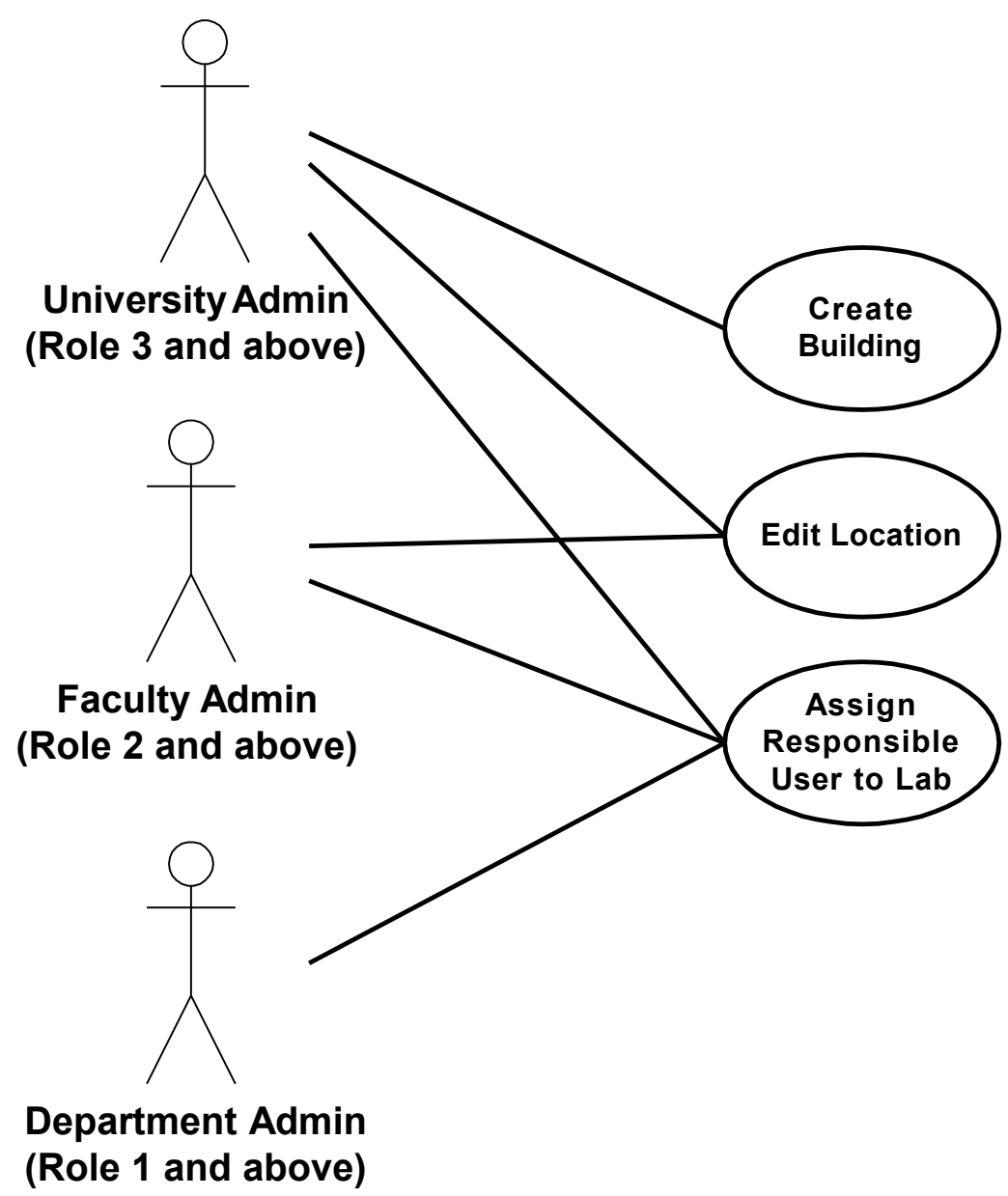

**Figure 12 –** Manage Locations – Use case diagram

The following paragraphs provide the detailed description for each use case.

*3.2.6.2.1 Edit Location*

**Role:** Faculty Admin Users (Role 2 and above);
**Pre-condition:** Authenticated session*, Edit Location* privileges;
**Steps:**

1. Admin selects the *Edit Location* option under the *Locations* tab (accessible from all system pages);
2. The system displays the *Edit Location* page;
3. The system prompts the admin to enter a location number;
4. The system displays the current status of the location including:
    a. Location Type (Research Lab, Office, or Room)
    b. Department Affiliation
    c. Faculty Affiliation (if current user is University Admin)
    d. User Responsible, if a Research Lab;
    e. Available or Unavailable (for construction or other reasons)
5. The Admin selects the parameter they wish to edit, from the above list
    a. Parameters a, b, c, and d offer a drop down menu to select the new status
    b. Parameter d, editing the user responsible, provides a link to the *Edit Lab Head* page.

Repeat step 4-5 until finished;

6. The user selects the *Submit* button;
7. The system logs the transaction;

**Post-conditions:**

1. The system displays the *Edit Locations* page;
2. A location has a new set of parameters;

**Exceptions:**

1. At any time system fails:
    a. The system displays an error message.
2. At any time, the Admin cancels by selecting the Cancel button:
    a. The system displays the *Edit Locations* page; and
    b. The system does not log transaction associated with the cancelled request;

*3.2.6.2.2 Assign Responsible User to Lab*

**Role:** Department Admin (Role 1 and above);
**Pre-condition:** Authenticated session;
**Steps:**

1. Admin selects the *Edit Lab Head* option under the *Locations* tab (accessible from all system pages);
2. The system displays the *Edit Lab Head* page;
3. The system prompts the admin to enter a location number;
4. Provided the dept admin entered a lab location in their department, the system displays the name and user id of the current person responsible, if any;
5. The admin enters the user id of the new person responsible for the lab;
6. The user selects the *Submit* button;
7. The system logs the transaction;

**Post-condition:**

1. The system displays the Edit Lab Head page;
2. The selected lab has a new person responsible for it

**Exceptions:**

1. At any time system fails:
    a. The system displays an error message.
2. The admin selects a location that is not in their permission to view;
    a. The system displays a message stating that the current location is not accessible by the admin
3. The admin selects a location that is not designated a research lab;
    a. The system displays a message stating that the selected location is not a research lab.

*3.2.6.2.3 Create Building*

**Role:** University Admin (Role 3)
**Pre-Conditions:** Authenticated session*, Create/Remove Location* privilege (rare privilege)
**Steps:**

1. Admin selects the *Create Building* option under the *Locations* tab (accessible from all system pages);
2. System prompts the user to enter their password before proceeding
3. System prompts for building information including:
   a. Address
   b. Name
4. The system searches if the building exists or not
   a. If the Building has not yet been entered, the admin selects the *Create Building* button, and the system saves an unfinished building with no locations or floors
   b. If the Building has begun being created but is not yet finished, the admin selects the Floor they wish to work on from a drop down menu of previously entered Floors
   c. If the Building has been *Completed* (see Step 10), and the current User is the supervisor who received the notification then:
      i. A Floor is selected from a drop down menu
      ii. A Location is selected from a drop down menu of all Locations on the selected Floor
      iii. The Location information is displayed
      iv. At any point in reviewing the information, the user (supervisor) has access to the *Commit Building* button, which commits the Building to the database, and allows Physical Assets to be placed within it's Locations
      v. The system logs the transaction;
5. System prompts for information on the next lowest floor including:
   a. Number/name
   b. Area
6. Admin selects either the *Create Floor* button, or the *Save and Exit* button.
   a. If *Create Floor* is selected, the system saves an unfinished floor with no locations, but connected to the unfinished building
   b. If *Save and Exit* is selected, the system registers all entered information, logs the transaction and returns to the system *Home Page*
7. System prompts for information on the first location of the floor including:
   a. number (required)
   b. area (required)
   c. type (Research Lab, Office, or Room – Optional)
   d. Faculty (optional)
   e. Dept (optional)
   f. Availability or Occupancy are not prompted. Availability is set to Unavailable by default.

8. Admin selects the *Create Location* button
9. System saves a new Location
10. System prompts for either a *New Location on this Floor, Move to the Next Floor, Save and Exit,* or *Building Complete*
    a. If *New Location on this Floor* is selected, Admin repeats Steps 9-12.
    b. If *Move to the Next Floor* is selected, the Admin repeats Steps 6-7.
    c. If *Save and Exit* is selected, the system registers all entered information
    d. If *Building Complete* is selected, a notification is sent to the user's supervisor to review the work.
11. The system logs the transaction;

**Post-Conditions:**

1. The System Displays the *Home Page*
2. Some information on the building is saved
3. If *Building Complete* was selected, a notification is sent out
4. If *Commit Building* was selected, a new set of locations exists to be used by the system

**Exceptions:**

1. At any time system fails:
    a. The system displays an error message.

**3.2.6.3 Functional Requirements**

*3.2.6.3.1 Edit Location*

**[F.Req.1]** The system shall allow the user to name the location
**[F.Req.2]** The system shall allow the user to determine the location type
**[F.Req.3]** The system shall allow the user to determine the location measurements
**[F.Req.4]** The system shall allow the user number locations
**[F.Req.5]** The system shall allow the user to view location map

*3.2.6.3.2 Assign Responsible User to Lab*

**[F.Req.6]** The system should allow the assigning of a lab to a TA
**[F.Req.7]** The system shall allow a lab to be assigned to many TAs
**[F.Req.8]** The system shall allow the lab to be shared according to a schedule

*3.2.6.3.3 Create Building*

**[F.Req.9]** The system shall allow the addition of a new building
**[F.Req.10]** The system shall allow the specification of the new building details

## 3.2.7 Bulk Entry (via CSV files)

The purpose of the bulk entry module is to allow the user to mass entry inventory assets through a CSV file.

### 3.2.7.1 Use Case Model

The following paragraphs provide the detailed description for each use case.

*3.2.7.1.1 Bulk Upload of CSV file*

This use-case shows the steps taken by the user and the UUIS application when a CVS is imported.

**Role:** User role 1 and above
**Precondition:** User is successfully authenticated. CSV file is properly formatted and created.
**Steps:**

1. User clicks on the button "BULK IMPORT" located at the top of every page.
2. System displays the bulk import page.
3. User clicks the "select a file" button on the file input form.
4. User selects from a radio buttons the type of asset he is importing (physical asset, software or location).
5. User clicks the "import into inventory" button.
6. System imports the selected file, validate its contents and put them into the inventory.
7. System displays a message of success (or failure) to the user.

**Post-Condition:** Uploaded items are now presents into the inventory.
**Exception:**

- Invalid file size.
- Invalid extension (only CVS are accepted)
- Contents of the file does not match the selected radio (physical asset, software or location)
- Database not functioning

### 3.2.7.2 Functional Requirements

*3.2.7.2.1 Bulk Upload of CSV file*

**[F.Req.1]** The bulk module shall support files with size less than 2mb.

**[F.Req.2]** The bulk module shall work atomically: everything or nothing is imported.

**[F.Req.3]** The bulk module shall be accessible: easy to navigate.
**[F.Req.4]** The bulk module shall be secure; only plain ASCII files with extension CVS will be accepted.
**[F.Req.5]** The bulk module shall sanitize every field in the file to prevent attacks.
**[F.Req.6]** The bulk module shall be qualitative; if an error happens with the import, the exact place of the error shall be told to the user.
**[F.Req.7]** The bulk module shall be stable; when an import is taking place, other users shall browse the inventory with no delays.
**[F.Req.8]** The bulk module shall support less than 10 concurrent imports.

## 3.3 Performance Requirements

**a) The number of terminals to be supported:**

Since the system will be deployed in a university, it should be able to support a high number of computers. There are typically 30,000 undergraduate students in a university located in a major city; 5,000 graduate students, and some 2,000 people composed of staff, researchers and professors. Our system should be ready to handle a user base of at least 50,000 users.

**b) The number of simultaneous users to be supported:**

Since our system will be a web based application, virtually every user with a valid user account, and who possesses a computer and internet connection in a remote location can access the application. Taking into account this fact, the system should be able to support at least 1,000 concurrent users.

**c) Amount and type of information to be handled:**

The system will store all the data into a MySQL Database. The data itself will be stored in a tabular form. Different tables will store different types of data. Examples of type of information to expect in the Database are strings, integers, and Boolean types.
The system should be able to handle large amounts of data that can reach terabytes of data. Taking into account the high number of users that will be using the system and its expected frequent use, we anticipate large amounts of data will be produced and stored in our Database.
In 95 % of cases, our system response time should be no more than 1 second. The system must compulsorily give a response to the user in the %5 of remaining cases in no more than 4 seconds. We expect the system to handle some 1,000 transaction per day most of which will be related to the module, Software management; this module will be the most used module on a regular basis by both students and faculty to get to know what software license is purchased by the university and is available for use by the university community.

Some modules like Manage Location will experience peak usage by faculty and students at the beginning of a study semester. As much as 2,000 users might search for a location on a daily basis during the 1st week of the semester; this number will fall after the 1st two weeks to an average 200 search per day.

## 3.4 Design Constraints

**Standards compliance:**

a) **Report format:**
The Reports written in preparation for the physical implementation of the system like the software Requirement Specification (SRS) or the Software Design Document (SDD) were written while keeping in mind the guidelines provided by the IEEE standards like IEEE830, Recommended Practice for Software Requirements Specifications, and IEEE1016, Recommended Practice for Software Design Descriptions.
These standards published by a prestigious organization, IEEE, provide us with a proven methodology on how to organize the different parts of the reports so that it is easier for readers to go look for the piece of information they look for fast.

b) **Data naming:**
The team has decided at the beginning of the project to adopt a naming convention. Since the group size is quite big, nine people, adopting the convention eliminate a great deal of the potential miscommunication that arises in situations like these. A naming convention that everyone in the team follows makes the task of working in a big group less troublesome.

## 3.5 Software System Attributes

a) **Reliability:**
We will determine that our system is reliable if it performs the different tasks required from it like authenticating the legitimate users of the system, validation of input and responding back with the right and expected output.

b) **Availability:**
An uptime guarantee of at least 99.9% should be guaranteed. A backup alternative for the system should be implemented too so the system would still be available for access even if the servers usually running the system go down for one reason or another.
Monthly maintenance tasks should be performed on the system to make sure it is running smoothly and that there are no serious issues that might affect its accessibility by the users.

c) **Security:**

The system uses many techniques to ensure data consistency and protection. Data is validated at both the client side and server side to make sure it conforms to the accepted format. Users are authenticated before accessing the system to determine their level of access and allow them do tasks associated with every access level.

d) **Maintainability:**
The system is developed using Object Oriented PHP. Object oriented programming allow programmers to reuse code more and write more readable code. The team decided to adopt the MVC, Model View Control model to build the application. Choosing a design pattern like MVC permits the separation between the presentations, business logic, and data layers. The application is more maintainable because dependencies and coupling between the different modules is minimized.

e) **Portability:**
The team has decided to use three technologies which are available on virtually any platform, Apache web server, PHP and MySQL. Apache, PHP and MySQL are popular open source software available for download for free. A system coded on Windows machines would be portable to a Linux server without modification to code. Only configuration details would differ from one platform to another. Other competing technologies like IIS and ASP.NET do not have this flexibility because they are Microsoft products.

## 3.6 Non-Functional Requirements

### 3.6.1 Multilingual support

**[N-F Req.1]** The system shall be bilingual (English / French)

**[N-F Req.2]** The system shall allow for upgrades to any number of languages

### 3.6.2 Backups

**Frequency**

**[N-F Req.3]** The system shall allow a user to adjust automatic backup frequency

**[N-F Req.4]** The system shall allow a user to perform a manual backup

**Storage Location**

**[N-F Req.5]** The system shall allow a user to specify a default location for remote backup and a default location for on-site backup

**Scale/scope**

**[N-F Req.6]** The system shall allow a user to specify the information to be backed up

### 3.6.4 Network architecture

**[N-F Req.7]** The system architecture shall be distributed

**[N-F Req.8]** The system shall use a broadband connection

### 3.6.6 Browser support

**[N-F Req.9]** The system shall function properly on top 10 browsers as determined by Google Trends

**[N-F Req.10]** The system shall allow mobile device browsing

### 3.6.7 Performance

**Query speed**

**[N-F Req.11]** The system shall display the results from database queries within 1 second

**Write speed**

**[N-F Req.12]** The system shall register new entries within 10 second of pressing a submit button

### 3.6.8 Users

**Number of users**

**[N-F Req.13]** The system shall support 50000 registered users

**Number of concurrent users**

**[N-F Req.14]** The system shall support 1000 concurrent users

### 3.6.9 Security

**Remote access**

**[N-F Req.15]** The system shall provide users with remote access

**User authentication**

**[N-F Req.16]** The system shall authenticate users through username and password login

**[N-F Req.17]** The system shall require passwords with 8 or more alphanumeric characters

**Permissions**

**[N-F Req.18]** The system shall enforce permissions on actions allowed to a user

**Denial of service attack**

**[N-F Req.19]** The system shall require a Captcha entry for each query after 5 queries in 30 seconds

**[N-F Req.20]** The system shall require a Captcha entry for each login attempt after 5 login attempts in 30 seconds

### 3.6.10 Device Support

**Barcode reader**

**[N-F Req.21]** The system shall provide a compatible interface to barcode reader model XYZ

### 3.6.11 System Crashes

**Recovery procedure**

**[N-F Req.22]** The system shall automatically alert the IT department upon system crash

**[N-F Req.23]** The system crash alert shall be an email and a SMS to the IT department head's cell phone

**[N-F Req.24]** The system shall display a page to users when the system crashes

**3.6.12 Error Handling**

**Database connection failure**

**[N-F Req.25]** The system shall, upon encountering a database error, display an error message to the user

**[N-F Req.26]** The system shall display "*Temporary service interruption*" message (hide database reasons)

**Number of users in excess of system capability**

**[N-F Req.27]** The system, upon experiencing a concurrent number of users exceeding its capability, shall display a "*Please wait, system is experiencing high load*"

**[N-F Req.28]** The system, if a user is unable to log in after 2 minutes due to high load, shall display a "*Please try again later*" message

**3.6.14 System Logs & Audits**

**[N-F Req.29]** The system shall log every Insert/Update/Delete transaction in a parallel database table

**[N-F Req.30]** The system shall log every time a user logs in and out

**[N-F Req.31]** The system shall allow users with administrator privileges to browse log entries according to user and date parameters

**[N-F Req.32]** The System shall allow users with administrator privileges to browse audit entries according to user, date and string-like parameter

# Appendix A – User Interface

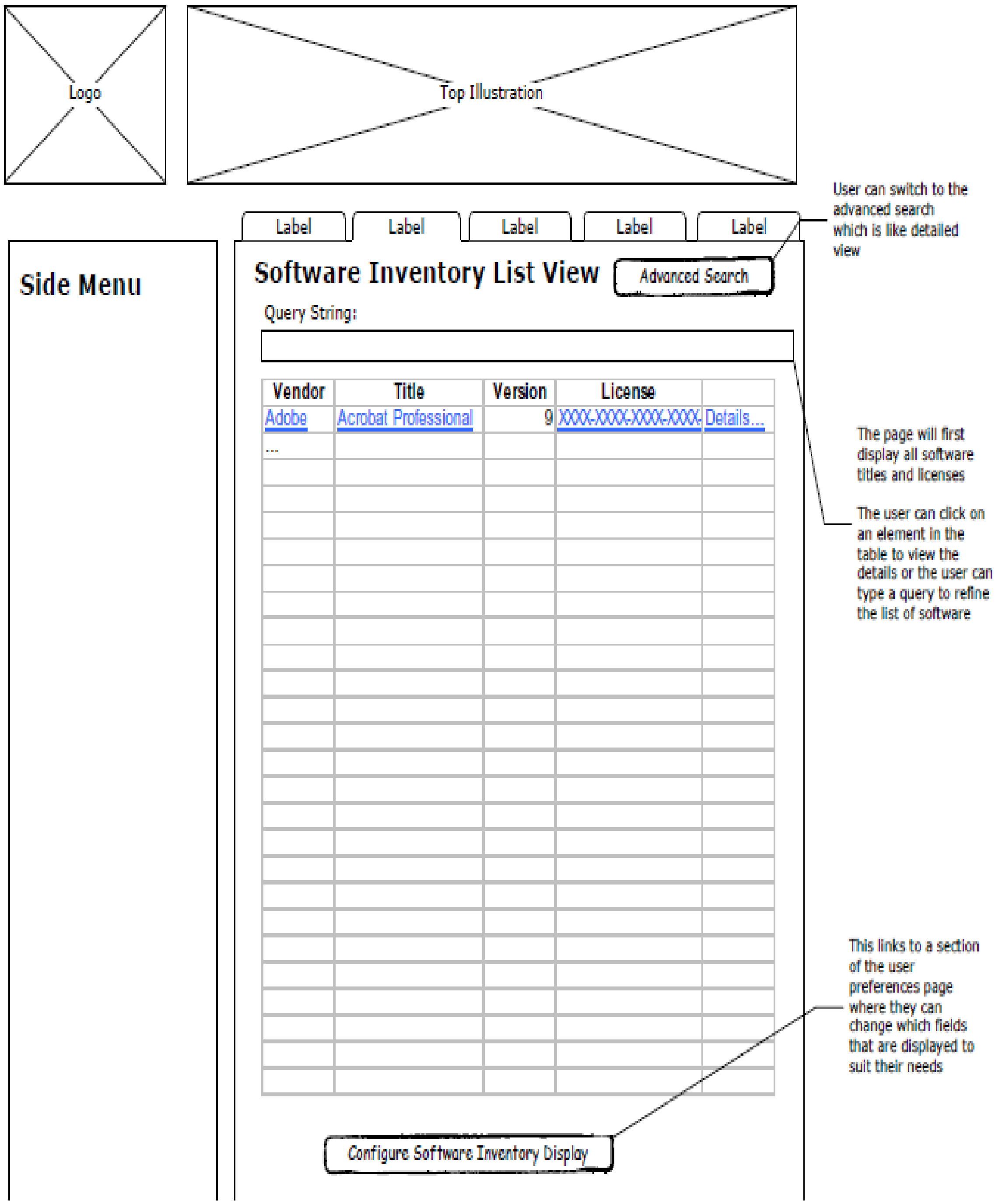

**Figure A. 1** - User Interface (1 of 12)

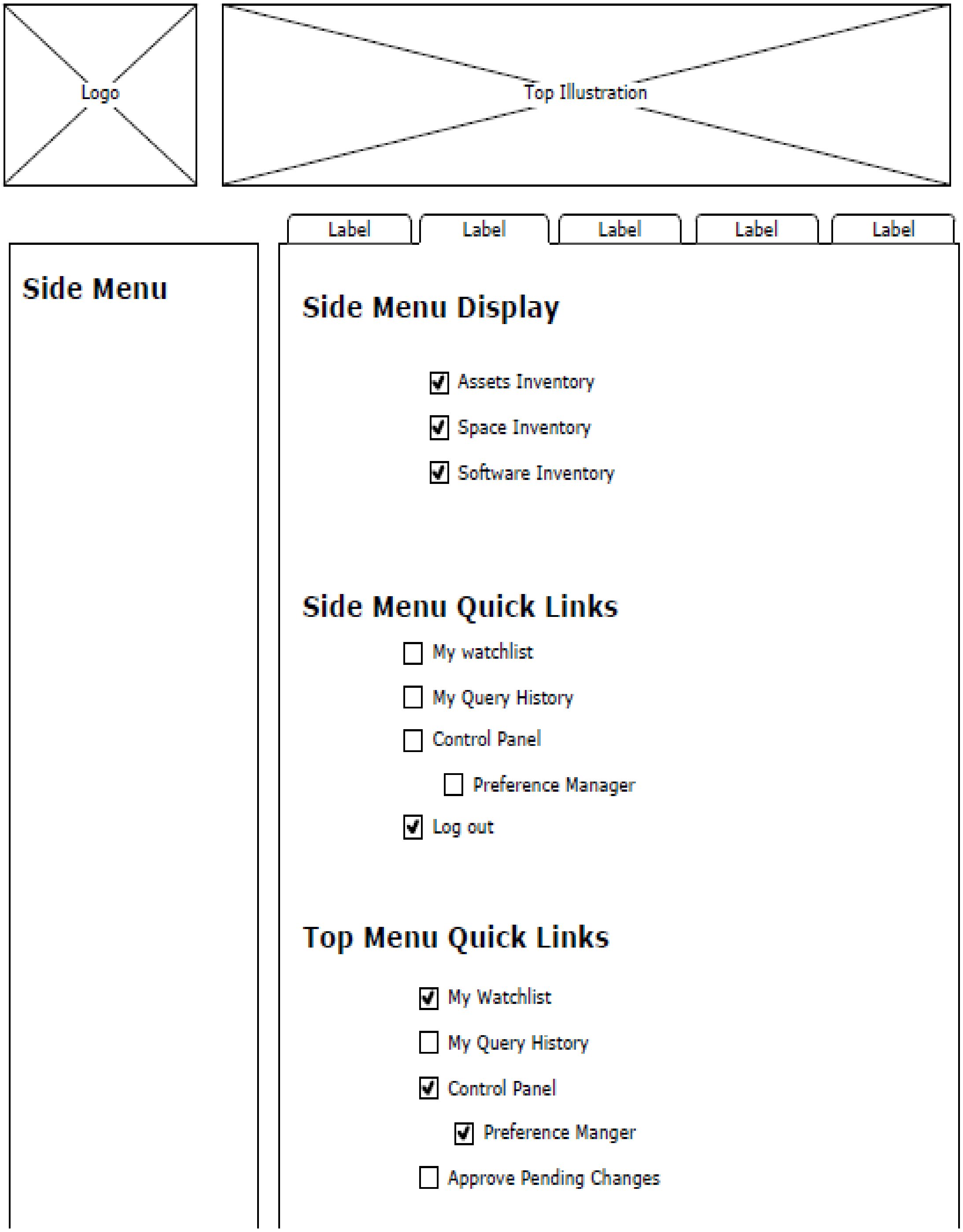

**Figure A. 2** - User Interface (2 of 12)

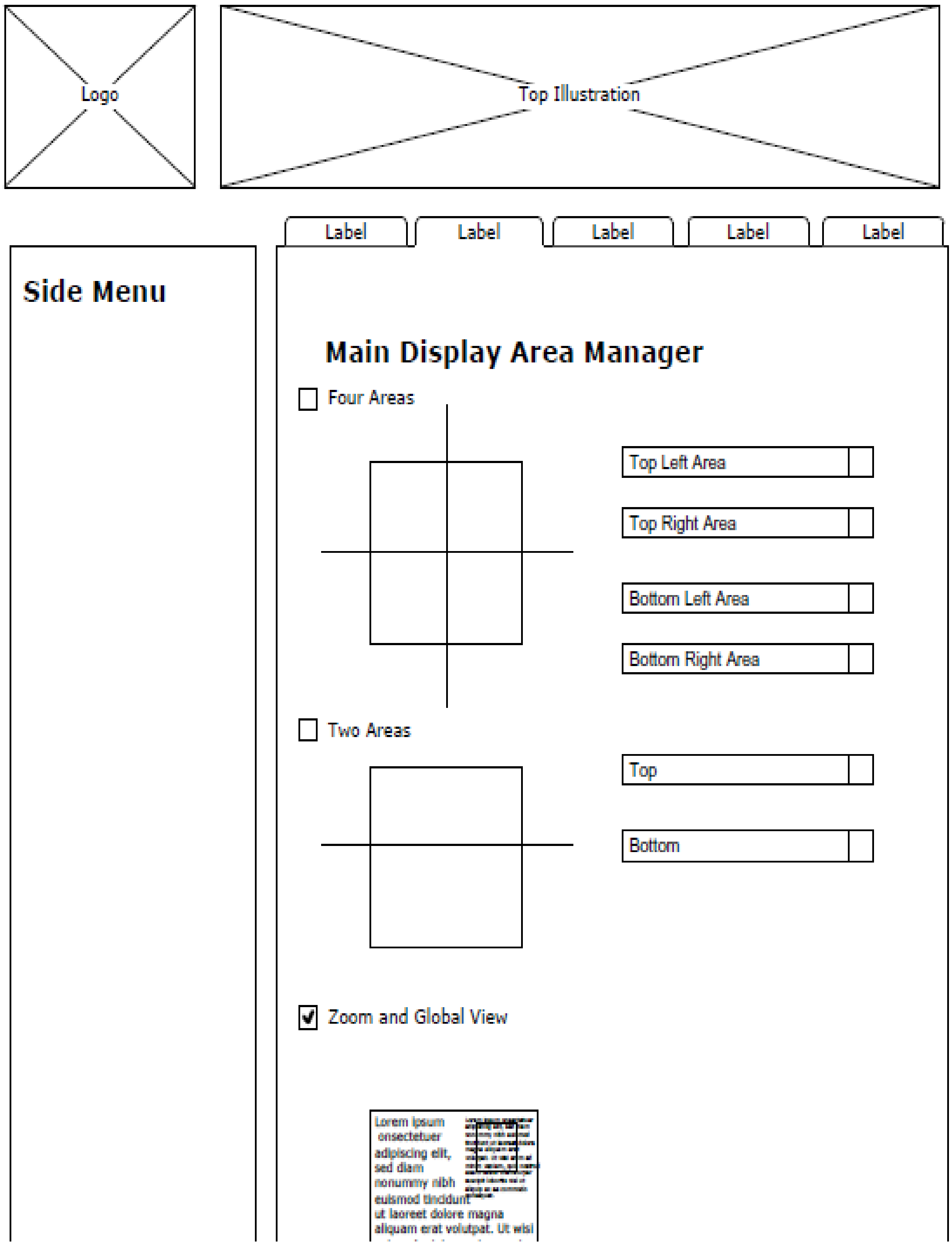

**Figure A. 3** - User Interface (3 of 12)

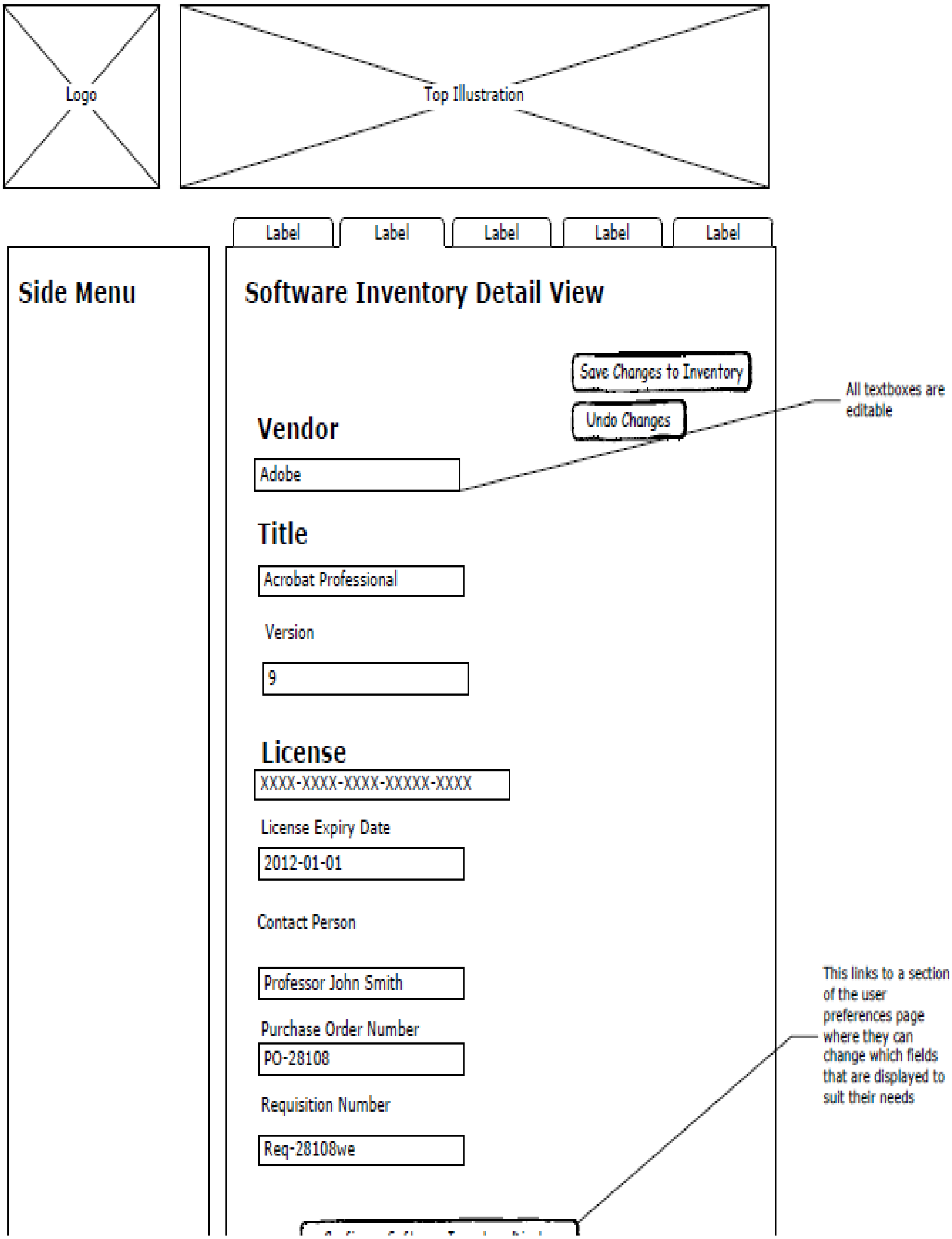

**Figure A. 4** - User Interface (4 of 12)

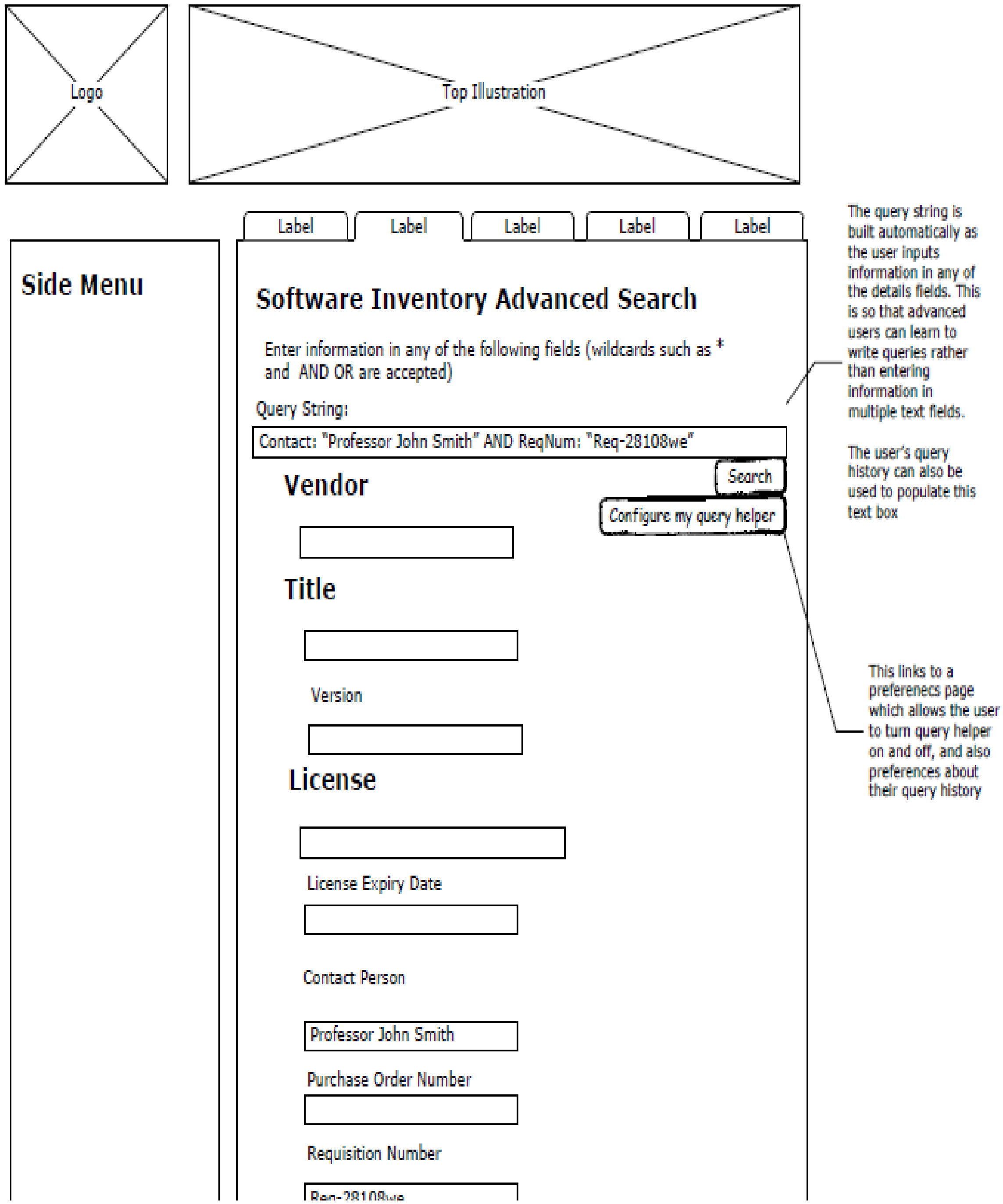

**Figure A. 5** - User Interface (5 of 12)

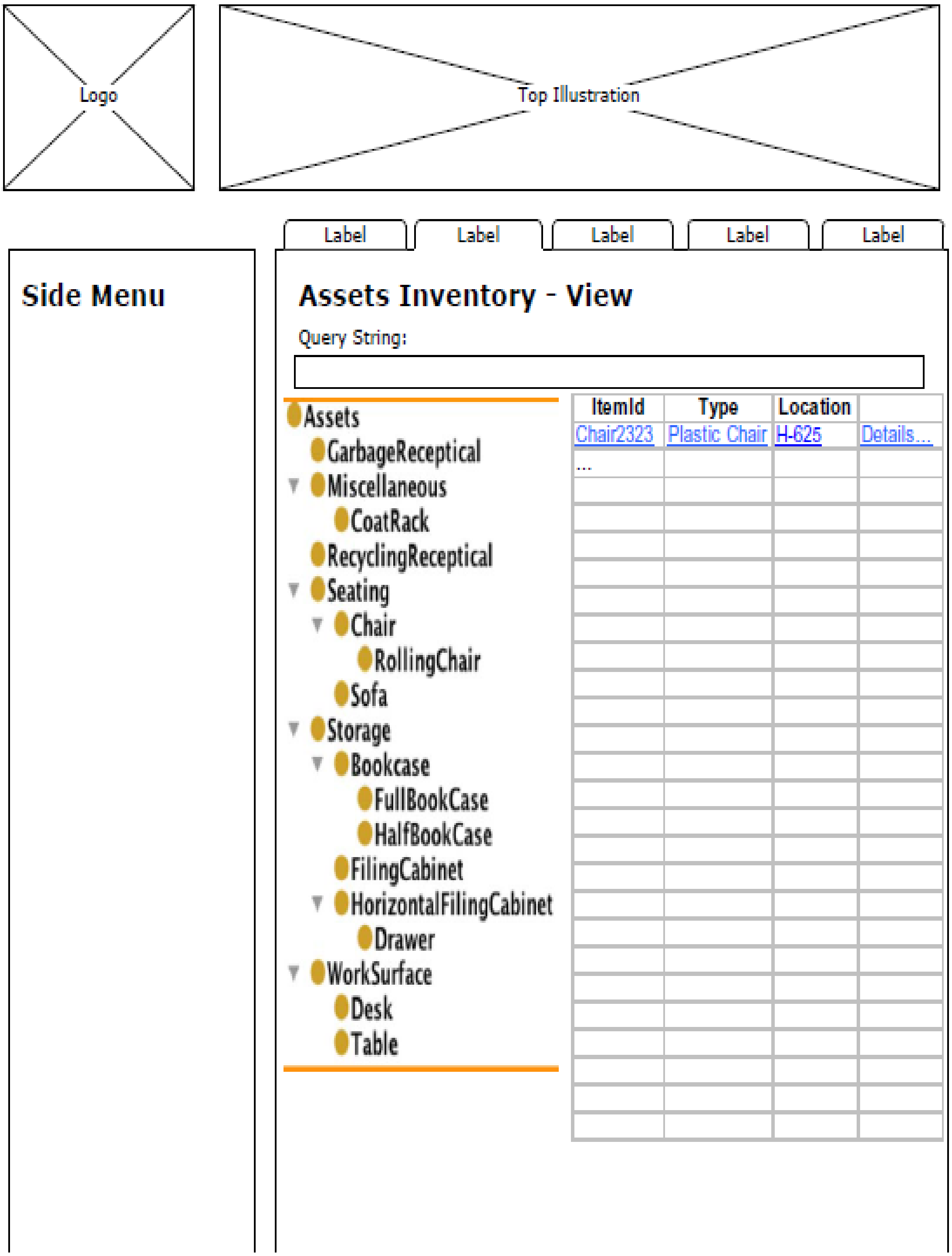

**Figure A. 6** - User Interface (6 of 12)

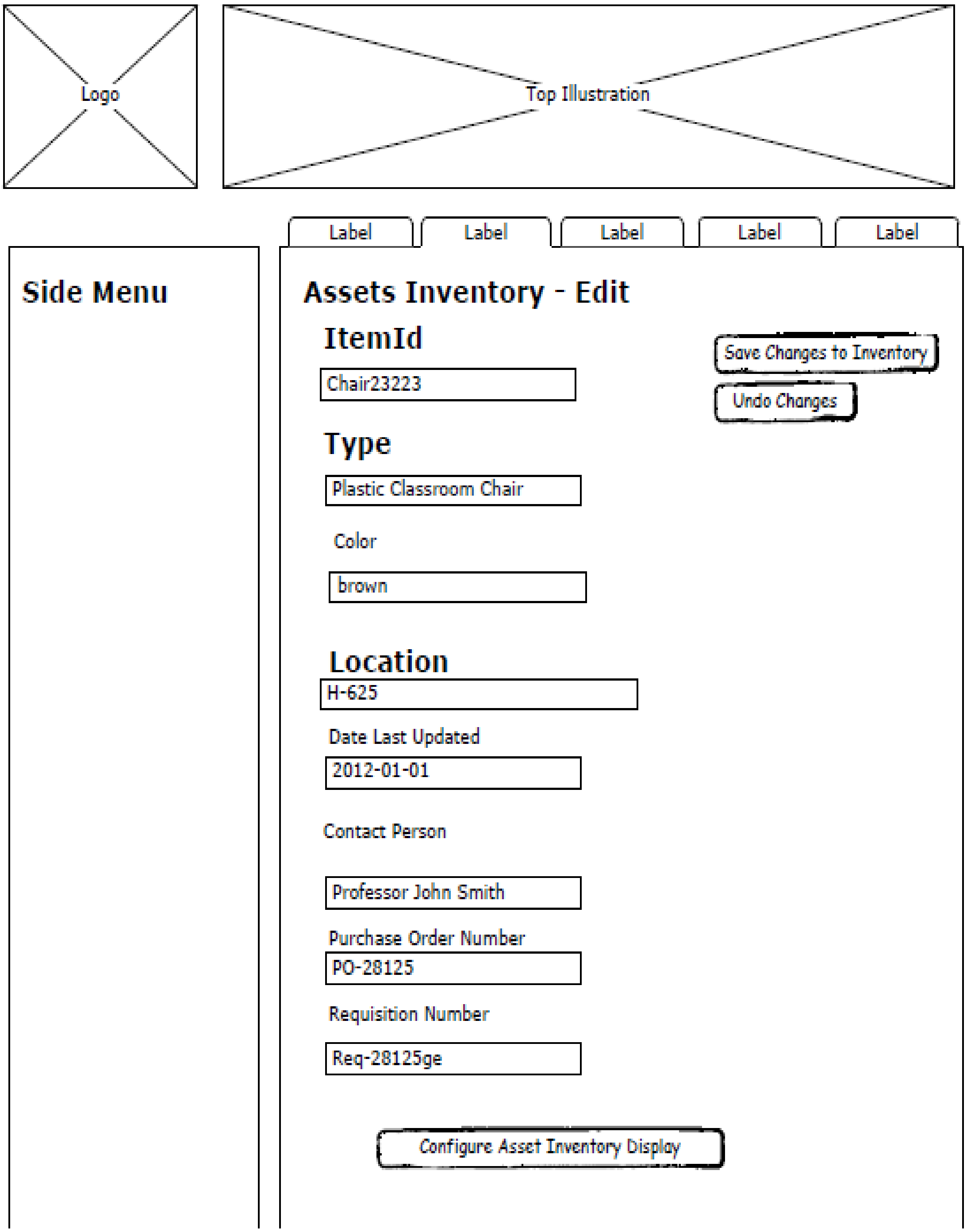

**Figure A. 7** - User Interface (7 of 12)

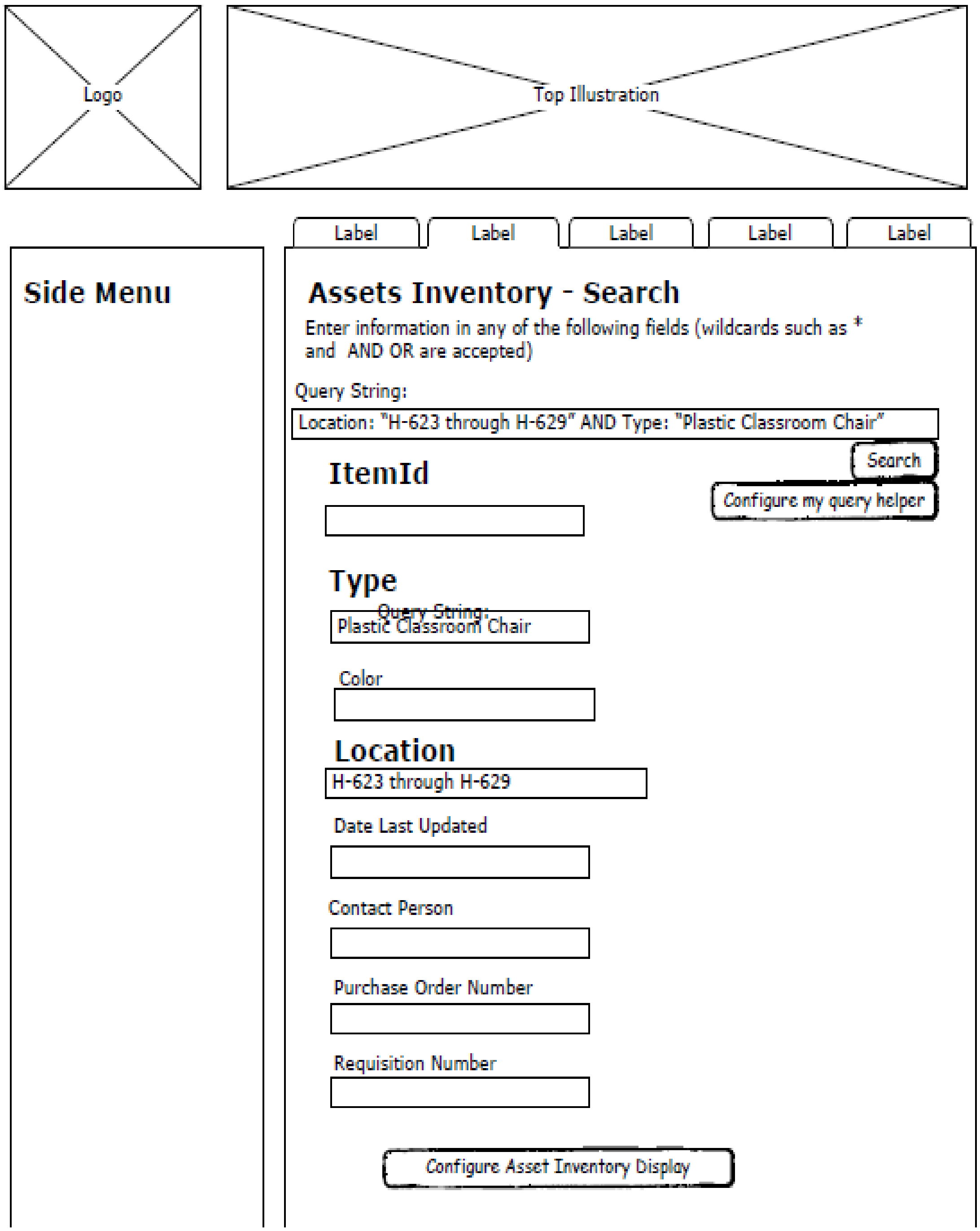

**Figure A. 8** - User Interface (8 of 12)

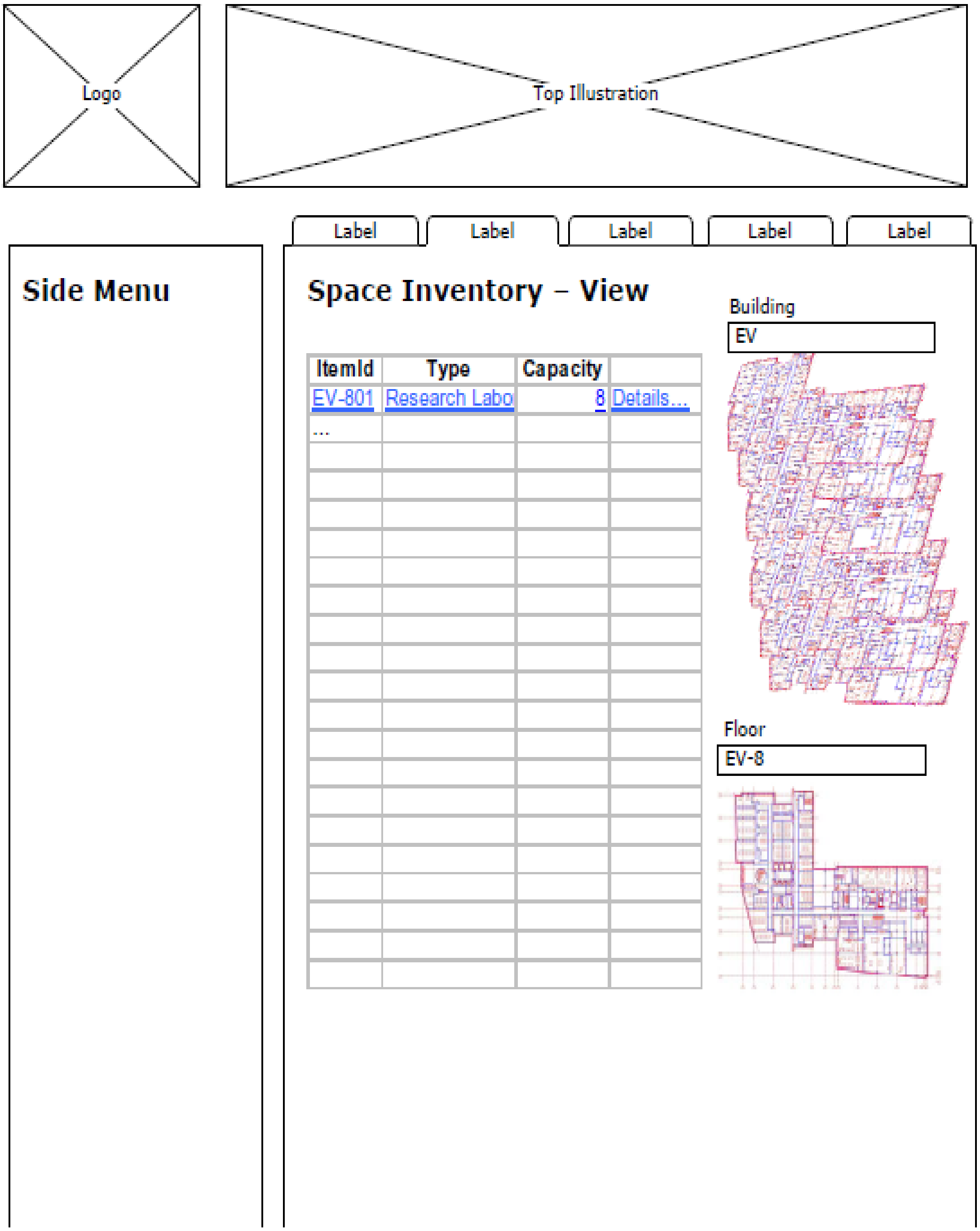

**Figure A. 9** - User Interface (9 of 12)

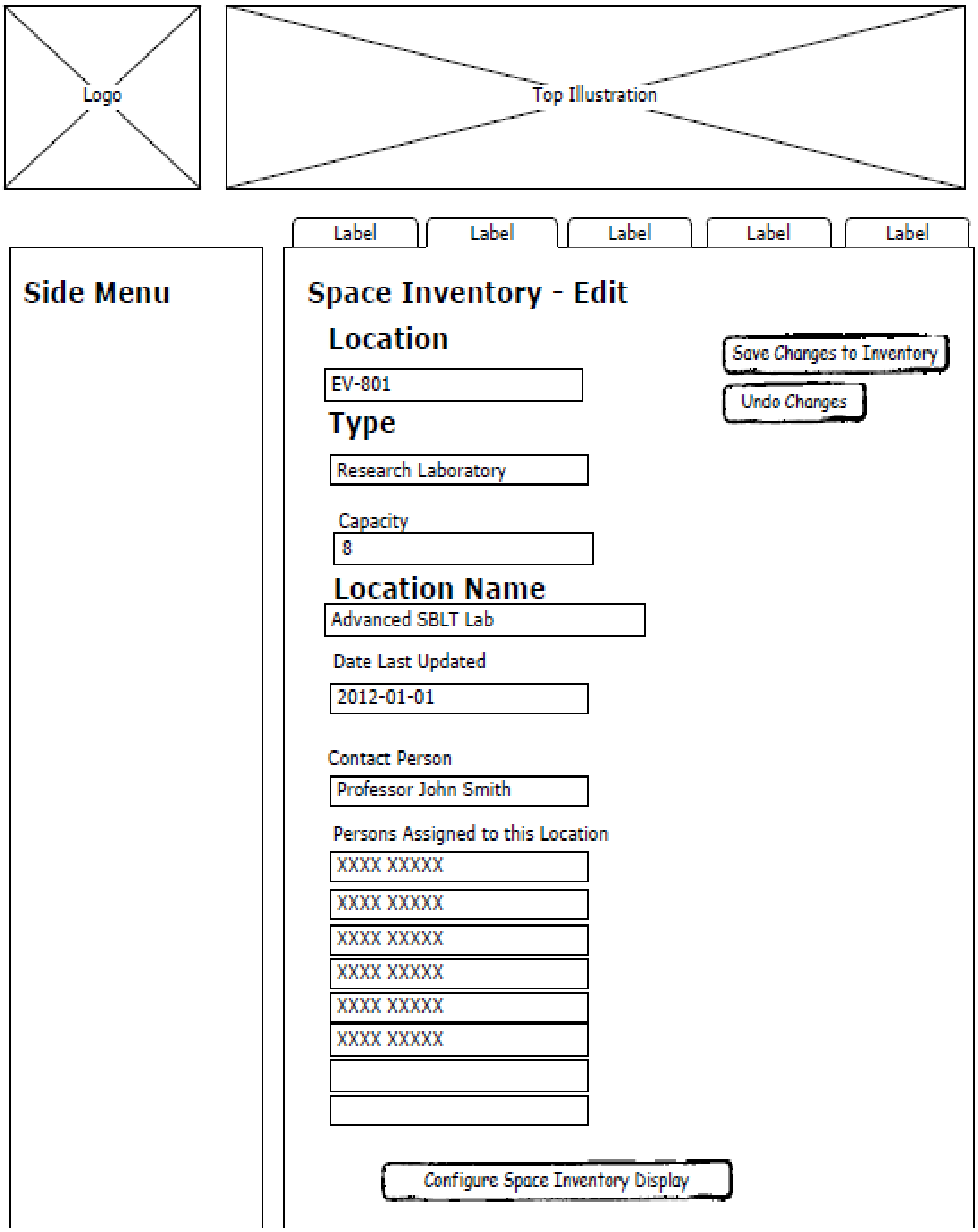

**Figure A. 10** - User Interface (10 of 12)

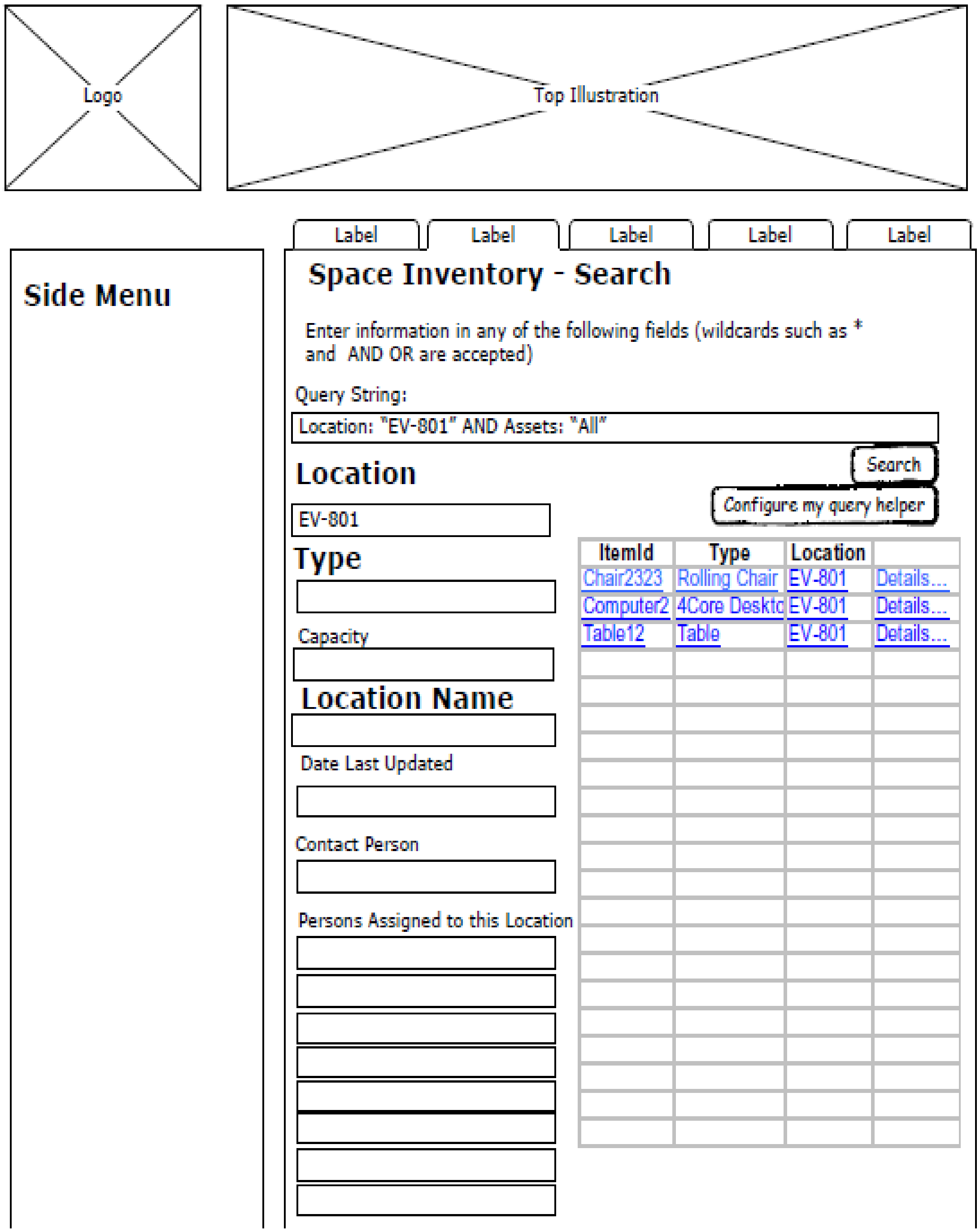

**Figure A. 11** - User Interface (11 of 12)

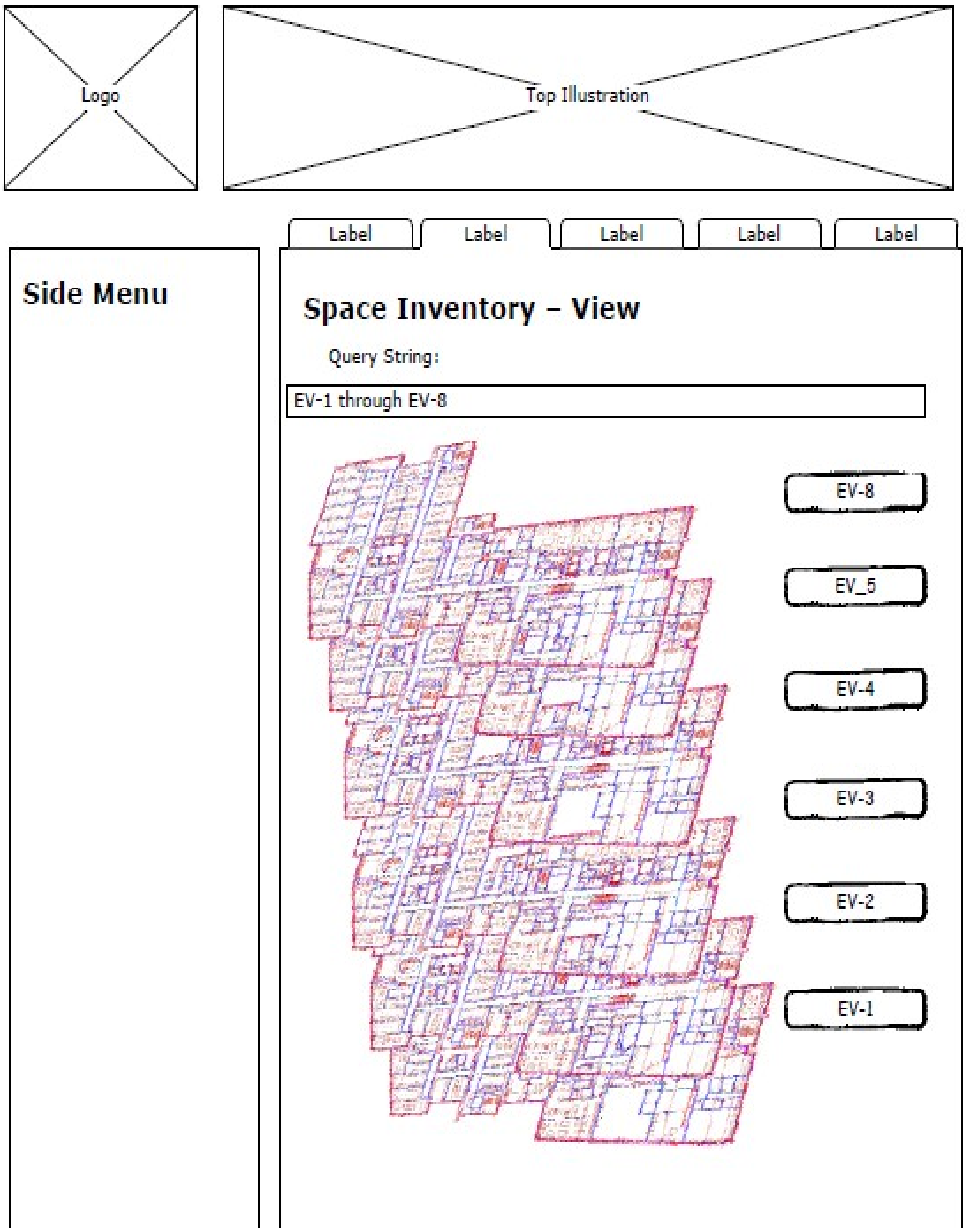

**Figure A. 12** - User Interface (12 of 12)